\documentclass[prd,aps,twocolumn,preprintnumbers, showpacs,
nofootinbib,superscriptaddress,notitlepage]{revtex4-1}

\usepackage[utf8]{inputenc}

\usepackage{amsthm,amsmath,amssymb}
\usepackage{mathrsfs}
\usepackage{graphicx}
\usepackage{footnote}

\usepackage{slashed}
\usepackage{hyperref}
\usepackage{physics}
\usepackage[utf8]{inputenc}
\usepackage{color}
\usepackage[normalem]{ulem}\usepackage{graphicx}

\begin{document}

\title{Nucleon Gluon Distribution Function from 2+1+1-Flavor Lattice QCD }

\author{Zhouyou Fan}
\affiliation{Department of Physics and Astronomy, Michigan State University, East Lansing, MI 48824}

\author{Rui Zhang}
\affiliation{Department of Physics and Astronomy, Michigan State University, East Lansing, MI 48824}
\affiliation{Department of Computational Mathematics,
  Science and Engineering, Michigan State University, East Lansing, MI 48824}

\author{Huey-Wen Lin}
\affiliation{Department of Physics and Astronomy, Michigan State University, East Lansing, MI 48824}
\affiliation{Department of Computational Mathematics,
  Science and Engineering, Michigan State University, East Lansing, MI 48824}

\preprint{MSUHEP-20-013}

%%%%%%%%%%%%%%%%%%%%%%%%%%%%%%%%%%%%%%%%%%%%%%%%%%%%%%%%%%%%%%%%%%%%%%%%%%%%%%%%
\begin{abstract}
The parton distribution functions (PDFs) provide process-independent information about the quarks and gluons inside hadrons. Although the gluon PDF can be obtained from a global fit to experimental data, it is not constrained well in the large-$x$ region. Theoretical gluon-PDF studies are much fewer than those of the quark PDFs. In this work, we present the first lattice-QCD results that access the $x$-dependence of the gluon unpolarized PDF of the nucleon. The lattice calculation is carried out with nucleon momenta up to 2.16~GeV, lattice spacing $a\approx0.12$~fm, and with valence pion masses of 310 and 690~MeV.
%\textcolor{blue}{Using these two pion masses, we attempt a naive extrapolation to estimate the gluon distribution at physical pion mass.}.
We use reduced Ioffe-time distributions to cancel the renormalization and implement a one-loop perturbative pseudo-PDF gluon matching. We neglect mixing of the gluon operator with the quark singlet sector. Our matrix-element results in coordinate space are consistent with those obtained from the global PDF fits of CT18 NNLO and NNPDF3.1 NNLO.
Our fitted gluon PDFs at both pion masses are consistent with global fits in the $x > 0.3$ region.
\end{abstract}

\maketitle

%%%%%%%%%%%%%%%%%%%%%%%%%%%%%%%%%%%%%%%%%%%%%%%%%%%%%%%%%%%%%%%%%%%%%%%%%%%%%%%%
\section{Introduction}

The unpolarized gluon parton distribution functions (PDFs) $g(x)$ and quark PDFs $q(x)$ are important inputs to many theory predictions for hadron colliders~\cite{Dulat:2015mca,Harland-Lang:2014zoa,Ball:2017nwa,Alekhin:2017kpj,Accardi:2016qay,Harland-Lang:2019pla,Bertone:2017bme,Manohar:2017eqh}. For example, both $g(x)$ and $q(x)$ contribute to the deep inelastic scattering (DIS) cross section, and $g(x)$ enters at leading order in jet production~\cite{Czakon:2013tha,Gauld:2015yia}. To calculate the cross section for these processes in $pp$ collisions, $g(x)$ needs to be known precisely. Although there are experimental data like top-quark pair production, which constrains $g(x)$ in the large-$x$ region, and charm production, which constrains $g(x)$ in the small-$x$ region, $g(x)$ is still experimentally the least known unpolarized PDF because the gluon does not couple to electromagnetic probes. The Electron-Ion Collider (EIC), which aims to understand the role of gluons in binding quarks and gluons into nucleons and nuclei, is at least in part intended to address this gap in our experimental knowledge~\cite{Accardi:2012qut}. In addition to experimental studies, the theoretical approaches to determining gluon structure by calculation are continually improving.

Lattice quantum chromodynamics (QCD) is a theoretical method that has full systematic control in calculating QCD quantities in the nonperturbative regime and can provide useful information for improving our knowledge of the gluon structure of the nucleon. However, there are much fewer lattice calculations of gluon structure than calculations of the nucleon isovector structure due to notorious noise-to-signal issues and complicated mixing in the renormalization. The few existing gluon-structure calculations were mostly done for the leading moments, such as the gluon momentum fraction~\cite{Yang:2018bft,Yang:2018nqn,Alexandrou:2020sml},
and nucleon gluon spin contribution~\cite{Alexandrou:2017oeh,Sufian:2016pex},
or at heavy quark mass, such as the gluon gravitational form factors of the nucleon and the pion~\cite{Shanahan:2018pib}. There has not been much effort to extract the $x$-dependent PDF for many decades.

In recent years, there has been an increasing number of calculations of $x$-dependent hadron structure in lattice QCD, following the proposal of Large-Momentum Effective Theory (LaMET)~\cite{Ji:2013dva,Ji:2014gla,Ji:2017rah}.
The LaMET method calculates on the lattice quasi-distribution functions, defined in terms of matrix elements of equal-time and spatially separated operators, and then takes the infinite-momentum limit to extract the lightcone distribution.
The quasi-PDF can be related to the $P_z$-independent lightcone PDF through a factorization theorem. The first part can be factorized into a perturbative matching coefficient, and the remaining part includes the corrections suppressed by the hadron momentum~\cite{Ji:2014gla}. This factorization can be calculated exactly in perturbation theory~\cite{Ma:2017pxb,Liu:2019urm}.
Alternative approaches to lightcone PDFs in lattice QCD are ``good lattice cross sections"~\cite{Ma:2017pxb,Bali:2017gfr,Bali:2018spj,Sufian:2019bol,Sufian:2020vzb} and the pseudo-PDF approach~\cite{Orginos:2017kos,Karpie:2017bzm,Karpie:2018zaz,Karpie:2019eiq,Joo:2019jct,Joo:2019bzr,Radyushkin:2018cvn,Zhang:2018ggy,Izubuchi:2018srq,Joo:2020spy,Bhat:2020ktg}.
There has been much progress made on the theoretical side since the first LaMET paper ~\cite{Xiong:2013bka,Ji:2015jwa,Ji:2015qla,Xiong:2015nua,Ji:2014hxa,Lin:2014yra,Monahan:2017hpu,Ji:2018hvs,Stewart:2017tvs,Constantinou:2017sej,Green:2017xeu,Izubuchi:2018srq,Xiong:2017jtn,Wang:2017qyg,Wang:2017eel,Xu:2018mpf,Chen:2016utp,Zhang:2017bzy,Ishikawa:2016znu,Chen:2016fxx,Ji:2017oey,Ishikawa:2017faj,Chen:2017mzz,Alexandrou:2017huk,Constantinou:2017sej,Green:2017xeu,Chen:2017mzz,Chen:2017mie,Lin:2017ani,Ishikawa:2019flg,Li:2016amo,Monahan:2016bvm,Radyushkin:2016hsy,Rossi:2017muf,Carlson:2017gpk,Ji:2017rah,Hobbs:2017xtq,Xu:2018eii,Jia:2018qee,Spanoudes:2018zya,Rossi:2018zkn,Liu:2018uuj,Ji:2018waw,Bhattacharya:2018zxi,Radyushkin:2018nbf,Zhang:2018diq,Li:2018tpe,Braun:2018brg,Ebert:2019tvc,Ji:2019ewn,Sufian:2020vzb,Shugert:2020tgq,Green:2020xco,Shanahan:2020zxr,Lin:2020ssv,Ji:2020ect,Ebert:2020gxr,Lin:2020ijm}, on the lattice-calculation side, nucleon and meson parton distribution functions (PDFs)~\cite{Lin:2013yra,Lin:2014zya,Chen:2016utp,Lin:2017ani,Alexandrou:2015rja,Alexandrou:2016jqi,Alexandrou:2017huk,Chen:2017mzz,Alexandrou:2018pbm,Chen:2018xof,Chen:2018fwa,Alexandrou:2018eet,Lin:2018qky,Fan:2018dxu,Liu:2018hxv,Wang:2019tgg,Lin:2019ocg,Chen:2019lcm,Chai:2020nxw,Bhattacharya:2020cen,Lin:2020ssv,Zhang:2020dkn,Bhat:2020ktg,Fan:2020nzz}; for more details, we refer readers to a few recent reviews~\cite{Detmold:2019ghl,Ji:2020ect,Lin:2020rut} and their references.
Although there are limitations of finite volume and relatively coarse lattice spacing, the latest nucleon isovector quark PDFs determined from lattice data at the physical point have shown reasonable agreement~\cite{Chen:2018xof,Lin:2018qky,Alexandrou:2018pbm} with phenomenological results from global fits to the experimental data~\cite{Dulat:2015mca,Ball:2017nwa,Harland-Lang:2014zoa,Nocera:2014gqa,Ethier:2017zbq}. However, the theoretical uncertainties and lattice artifacts need to be carefully studied to obtain fully reliable results. The latest efforts include an analysis of finite-volume systematics~\cite{Lin:2019ocg} and exploration of machine learning~\cite{Zhang:2019qiq,Zhang:2020dkn}.

The unpolarized gluon PDF is defined as the Fourier transform of the lightcone correlation of the nucleon,
\begin{align}
 g(x,\mu^2) &=\int\frac{\textrm{d}\xi^-}{\pi x}e^{-ix\xi^-P^+}  \nonumber \\
 & \times\langle P|F^+_{\mu}(\xi^-)U(\xi^{-},0)F^{\mu +}(0)|P\rangle,
\end{align}
where $\xi^\pm=\frac{1}{2}(\xi^0\pm\xi^3)$ are the spacetime coordinates along the lightcone direction, the nucleon momentum $P_\mu=(P_0,0,0,P_z)$ and $P^\pm=\frac{1}{2}(P_0\pm P_z)$, $|P\rangle$ is the hadron state with momentum $P$ with normalization $\langle P|P\rangle=1$, $\mu^2$ is the $\overline{\text{MS}}$ renormalization scale, $U(\xi^{-},0)=\mathcal{P}\exp(-ig\int^{\xi^-}_0d\eta^-A^+(\eta^-))$ is the lightcone Wilson link from $\xi^+$ to 0 with $A^+$ as the gluon potential in the adjoint representation, $F_{\mu\nu}=T^aG^a_{\mu\nu}=T^a(\partial_{\mu}A^a_{\nu}-\partial_{\nu}A^a_{\mu}-gf^{abc}A^b_{\mu}A^c_{\nu})$ is the gluon field tensor in the adjoint representation, $g$ is the coupling constant of the strong interaction, and $f^{abc}$ are the structure constants of SU(3).
A straightforward way to calculate the gluon PDFs directly on the lattices would be to use LaMET.
This was attempted when the first unpolarized gluon quasi-PDF matrix element was calculated in Ref.~\cite{Fan:2018dxu}; however, not all the operators used in the calculation can be multiplicatively renormalized, and the largest momentum used is only 1.3~GeV, where noise already dominated the signal.
Since then, there has been new development of the general factorization formula for the quasi-PDFs with the corresponding one-loop matching kernel calculated in Refs.~\cite{Zhang:2018diq,Wang:2019tgg} for the unpolarized and polarized gluon quasi-PDFs. In their papers, the authors also provide the multiplicatively renormalizable unpolarized and polarized gluon operators and the corresponding renormalization condition that would allow us to match the nonperturbatively renormalized gluon quasi-PDFs to the lightcone PDFs from lattice simulations.
However, calculating the gluon renormalization nonperturbatively suffers worse signal-to-noise than the corresponding nucleon calculation, making it harder to apply the strategies proposed in Refs.~\cite{Zhang:2018diq,Wang:2019tgg}.

In this work, we adapt the pseudo-PDF approach. It uses Ioffe-time distributions (ITDs) which are functions of Ioffe time $\nu = zP_z$ and the squared spacetime interval $z^2$. The pseudo-PDF approach uses ``reduced'' ITDs~\cite{Orginos:2017kos}, where the renormalization constants are canceled by taking ratios of the matrix element with corresponding that of the nucleon at rest. This approach is compared with the nonperturbative renormalization strategy in Refs.~\cite{Gao:2020ito,Fan:2020nzz}. This ratio not only removes Wilson-line--related UV divergences but also part of the higher-twist contamination. Recently, a methodology for determining the gluon PDFs within the pseudo-PDF approach was proposed in Ref.~\cite{Balitsky:2019krf}, allowing us to explore the gluon PDF without facing the noisy nonperturbative renormalization.
There have been a number of successful pseudo-PDF calculations of nucleon and pion PDFs.
Table~\ref{table-pseudopara} shows a summary of the lattice parameters used in calculations of $x$-dependent PDFs using the pseudo-PDF method.

\begin{table*}[!htbp]
\centering
\begin{tabular}{|c|c|c|c|c|c|c|c|c|c|}
\hline
  Reference  & PDFs & Sea quarks &   Valence quarks &  $P_\text{max}$ (GeV) &  $a$ (fm) &   $M_\pi$ (MeV) & $M_\pi L$ & $\mu$ (GeV) \\ % & $z^2$ \\
\hline
JLab/W\&M'17~\cite{Orginos:2017kos} & nucleon valence PDF  & clover &   clover &  2.5  & 0.09 & 601 & 8.8 &  $\{1,2\}$ \\ % & N/A\\
\hline
JLab/W\&M'19-1~\cite{Joo:2019jct} & nucleon valence PDF  & clover & clover & 2.44 & 0.094--0.127   & 390--415 & 4.5--8.6 &  2 \\ %&  & $4e^{-2\gamma_E-1}\mu^{-2}$ \\
\hline
JLab/W\&M'19-2~\cite{Joo:2019bzr} & pion valence PDF & clover & clover & 1.22 & 0.127   & 415 & 6.4--8.6 &  2 \\ %&  & N/A   \\
\hline
JLab/W\&M'20~\cite{Joo:2020spy} & nucleon valence PDF & clover & clover &  3.29 &  0.09  &  172--358 &  4.0--5.2  &  2 \\ %&  & N/A\\
\hline
ETMC'20~\cite{Bhat:2020ktg} & nucleon valence PDF & twisted-mass  & twisted-mass &  1.38 &  0.09  &  130 &  2.8  &   2 \\ %&  & $4e^{-2\gamma_E-1}\mu^{-2}$ \\
\hline
MSULat'20 (this work) & gluon PDF & clover & HISQ & 2.16 &  0.12  & 310--680 &  4.5--10  & 2 \\ %&  & $exp(-\gamma_e-1/2)/\mu$\\
\hline
\end{tabular}
\caption{The lattice parameters used in pseudo-PDF calculations of $x$-dependent PDFs.
Most of these works set the $\overline{\text{MS}}$ renormalization scale $\mu$ to 2~GeV (JLab/W\&M'17~\cite{Orginos:2017kos} also look at $\mu$ at 1~GeV). $z^2\mu^2$ is chosen such that the logarithm term in Eq.~\ref{matching-evolve} vanishes.
}
\label{table-pseudopara}
\end{table*}

The structure of this paper is organized as follows.
In Sec.~\ref{sec:set_up}, we present the numerical setup of lattice simulation and discuss the procedure to extract bare gluon ground-state matrix elements from the lattice data.
Section~\ref{sec:results} shows the numerical details to extract the physical pion mass unpolarized gluon distribution from the reduced Ioffe time pseudo-distribution and compares our results with the phenomenological global-fit gluon PDFs.
We summarize the final result and discuss future planned calculation in Sec.~\ref{sec:summary}.

%%%%%%%%%%%%%%%%%%%%%%%%%%%%%%%%%%%%%%%%%%%%%%%%%%%%%%%%%%%%%%%%%%%%%%%%%%%%%%%%
\section{Lattice Setup and Matrix Elements}\label{sec:set_up}

This calculation is carried out using the $N_f = 2+1+1$ highly improved staggered
quarks (HISQ)~\cite{Follana:2006rc} lattices generated by the MILC collaboration~\cite{Bazavov:2012xda} with spacetime dimensions $L^3\times T=24^3\times 64$, lattice spacing $a=0.1207(11)$~fm, and $M_\pi^\text{sea} \approx 310$~MeV. We apply 1 step of hypercubic (HYP) smearing~\cite{Hasenfratz:2001hp} to reduce short-distance noise. The Wilson-clover fermions are used in the valence sector where the valence-quark masses is tuned to reproduce the lightest light and strange sea pseudoscalar meson masses (which correspond to pion masses 310 and 690~MeV, respectively), as done by PNDME collaboration~\cite{Rajan:2017lxk,Bhattacharya:2015wna,Bhattacharya:2015esa,Bhattacharya:2013ehc}.
As demonstrated by PNDME and through our own calculation, we do not observe any exceptional configurations in our calculations caused by the mixed-action setup. Since our strange and light pion masses are tuned to match the corresponding sea values, we do not anticipate lattice artifacts other than potential $O(a)$ effects. Since this is at the same level as typical corrections to LaMET-type operators~\cite{Chen:2017mie}, it requires no special treatment. Such effects will be studied in future work.
%Our strange-quark mass and light-quark mass for the valence quark, and for the sea quark are chosen to be almost the same - i.e. the valence-sea quark mass gap is very small.

We first calculate two-point nucleon ($N$) correlator
\begin{align}\label{}
C_N^\text{2pt}(P_z;t)&=
 \langle 0|\Gamma\int d^3y\, e^{-iyP_z}\chi(\vec y,t)\chi(\vec 0,0)|0\rangle,
\end{align}
where $P_z$ is the boosted nucleon momentum along the spatial $z$-direction,
the nucleon interpolation operator $\chi$ is $\epsilon^{lmn}[{u(y)^l}^Ti\gamma_4\gamma_2\gamma_5 d^m(y)]u^n(y)$ (where $\{l,m,n\}$ are color indices, $u(y)$ and $d(y)$ are the quark operators),
projection operator is $\Gamma=\frac{1}{2}(1+\gamma_4)$,
and $t$ is lattice Euclidean time.
Gaussian momentum smearing~\cite{Bali:2016lva} is used for the quark field,
\begin{align}\label{eq:momsmear}
S_\text{mom}\Psi(x)=\frac{1}{1+6\alpha}\big(\Psi(x)+\alpha\sum_jU_j(x)e^{ik\hat{e}_j}\Psi(x+\hat{e}_j)\big),
\end{align}
where $k$ is the momentum-smearing parameter and $\alpha$ is the Gaussian smearing parameter. In our calculation, we choose $k=2.9$, $\alpha=3$ with 60 iterations to help us getting a better signal at a higher boost nucleon momentum. These parameters are chosen after carefully scanning a wide parameter space to best overlap with our desired boost momenta. We use 898 lattices in total and calculate 32 sources per configuration for a total 28,735 measurements.
In the previous gluon-PDF work~\cite{Fan:2018dxu}, the nucleon two-point function was calculated with overlap fermions using all timeslices with a 2-2-2 $Z^3$ grid source and low-mode substitution~\cite{Li:2010pw,Gong:2013vja}, which has 8 times more statistics and best signal at zero nucleon momentum.
Even though the number of measurements in this work is smaller than the previous work, we see significant improvement in the signal-to-noise at large boost momenta with our momentum smearing, which allow us to extend our calculation to momenta as high as 2.16~GeV. We studied the $(ap)^n$ discretization effects on the nucleon two-point correlators using ensembles of different lattice spacing $a\approx 0.6, 0.9, 0.12$~fm, and the results indicate that these effects are not significant on the two-point correlators. We anticipate the discretization effects to be small in our calculation, based on the observation in the two-point correlators; a study using multiple lattice spacings for the gluon three-point correlators will be needed for future precision calculations.

\begin{figure*}[htbp]
\centering
	\centering
	\includegraphics[width=0.42\linewidth]{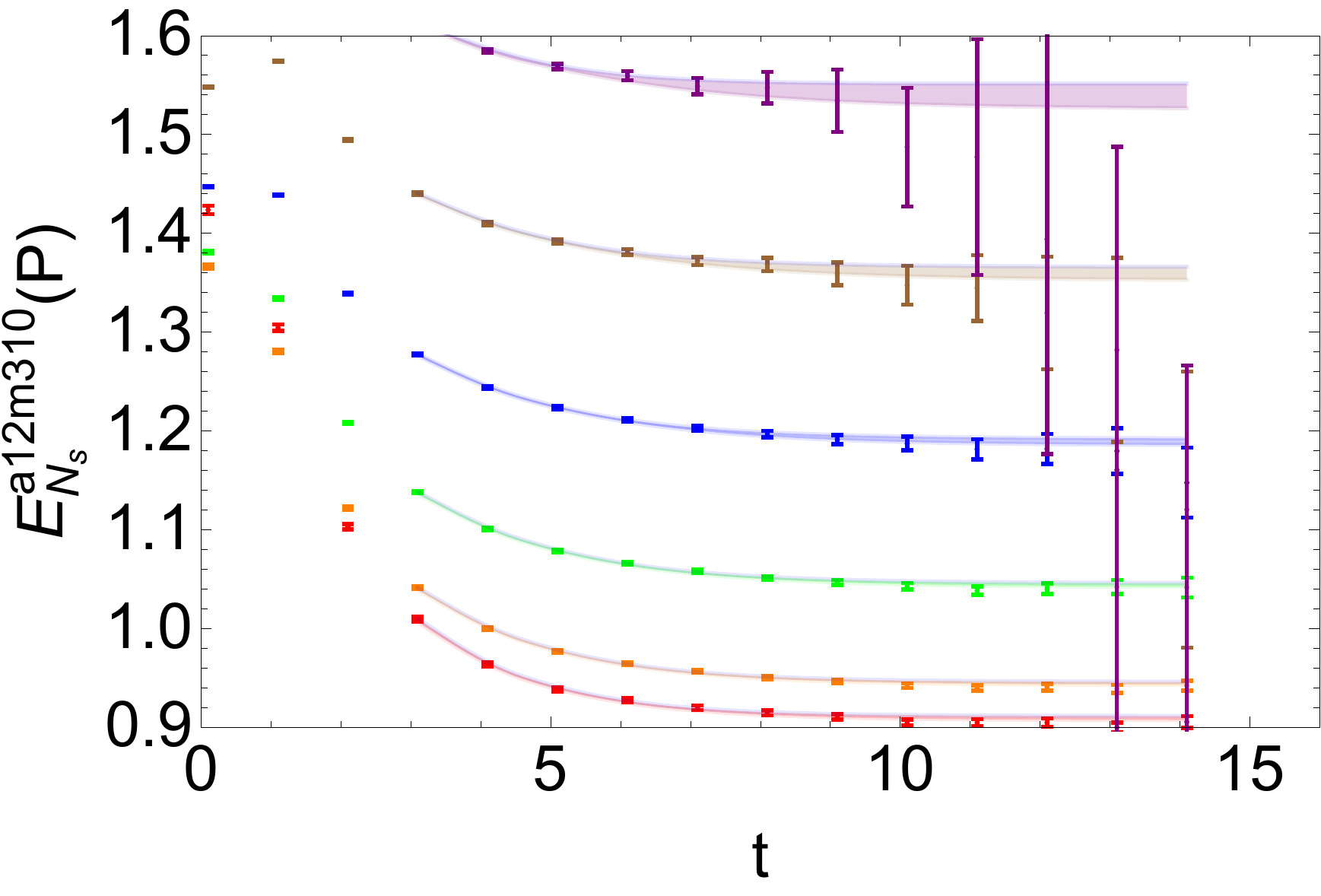}
    \includegraphics[width=0.42\linewidth]{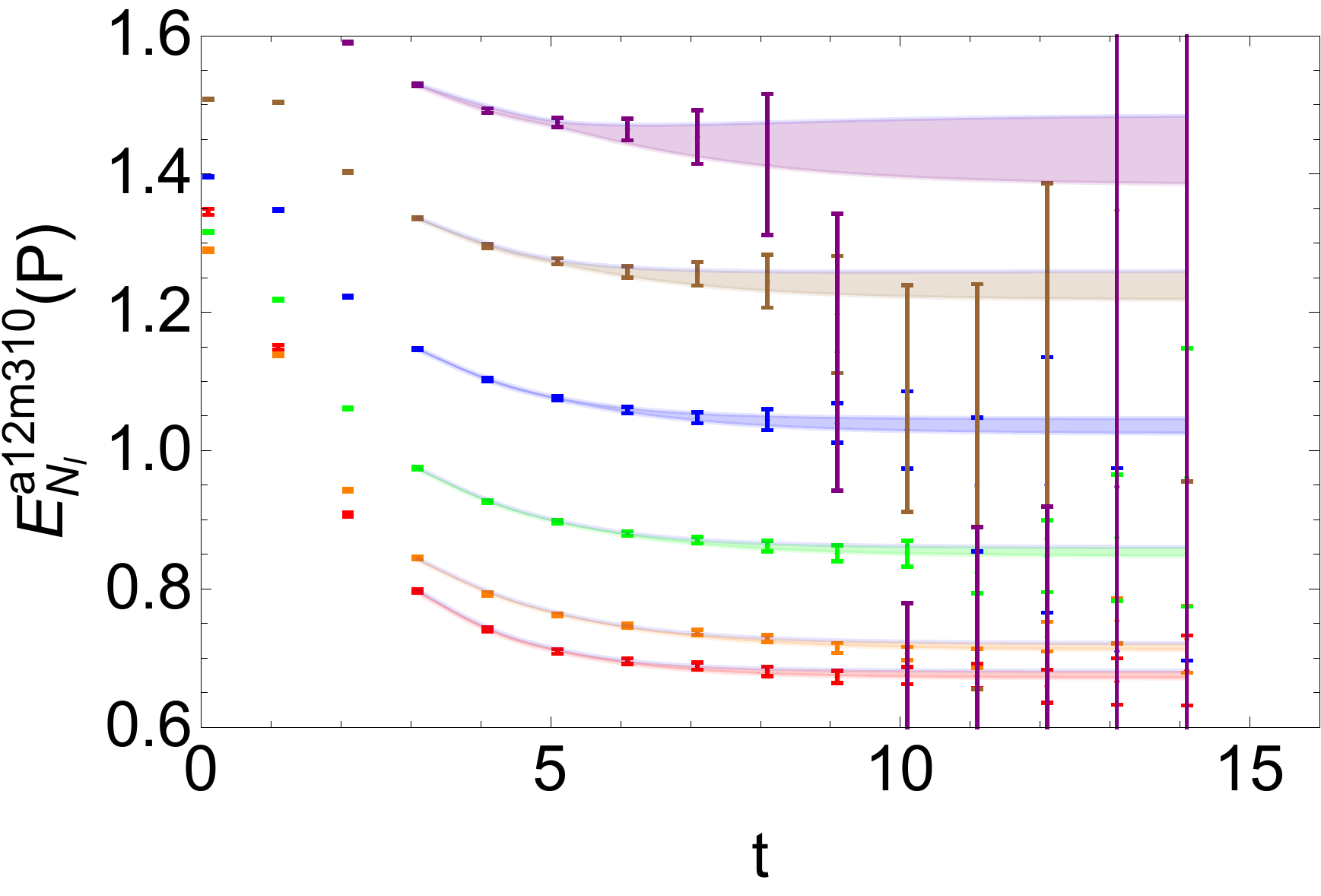}
\caption{Nucleon effective-mass plots for $M_\pi \approx 690$~MeV (left) and $M_\pi \approx 310$~MeV (right) at $z=0$, $P_z=[0,5]\times\frac{2\pi}{L}$ on the a12m310 ensemble. The bands are reconstructed from the two-state fitted parameters of two-point correlators. The momentum $P_z=5\frac{2\pi}{L}$ is the largest momentum we used, and it is the noisiest data set.}
\label{fig:effectivemass}
\end{figure*}

\begin{figure*}[htbp]
\centering
	\centering
	\includegraphics[width=0.42\linewidth]{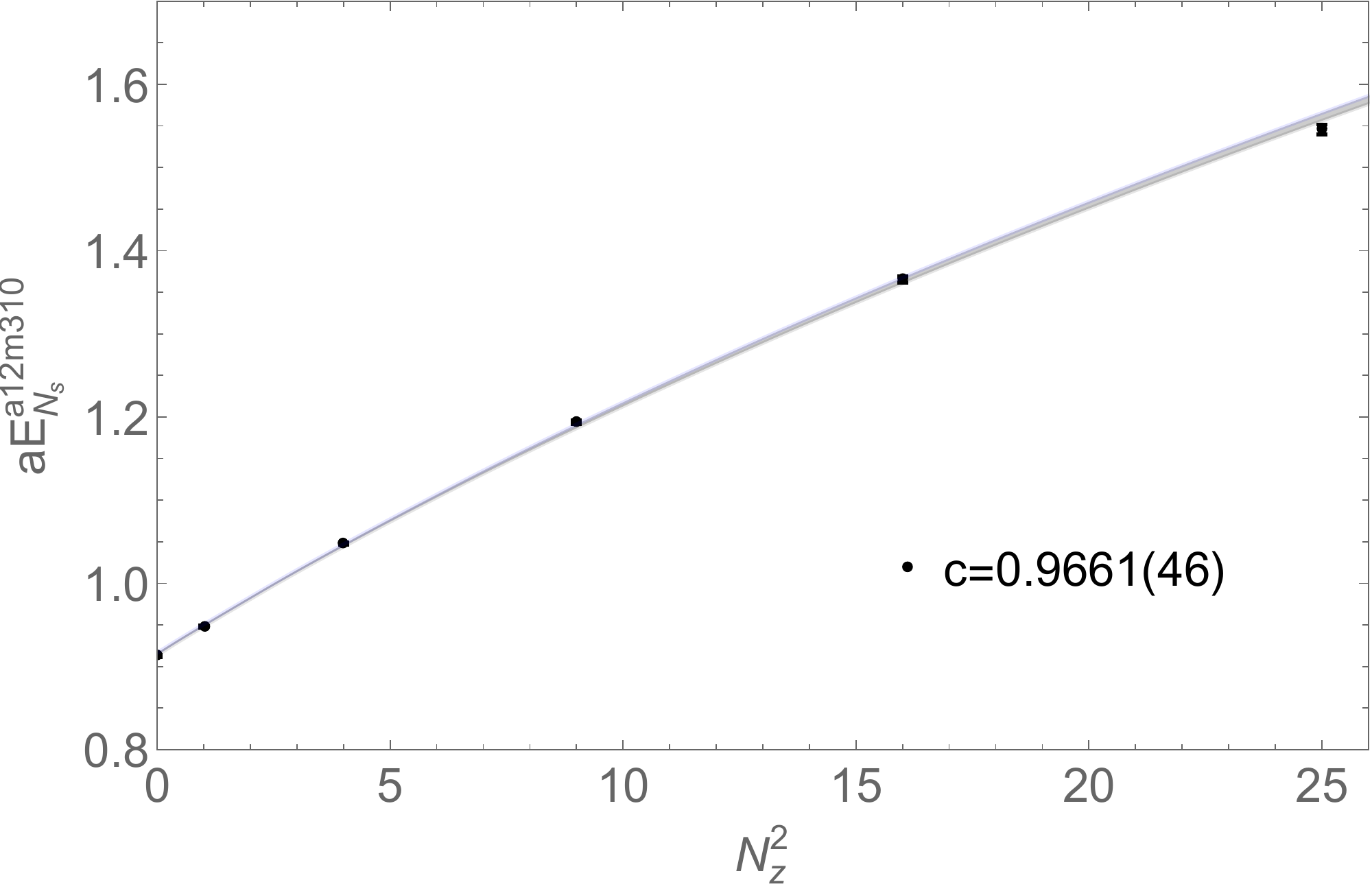}
    \includegraphics[width=0.42\linewidth]{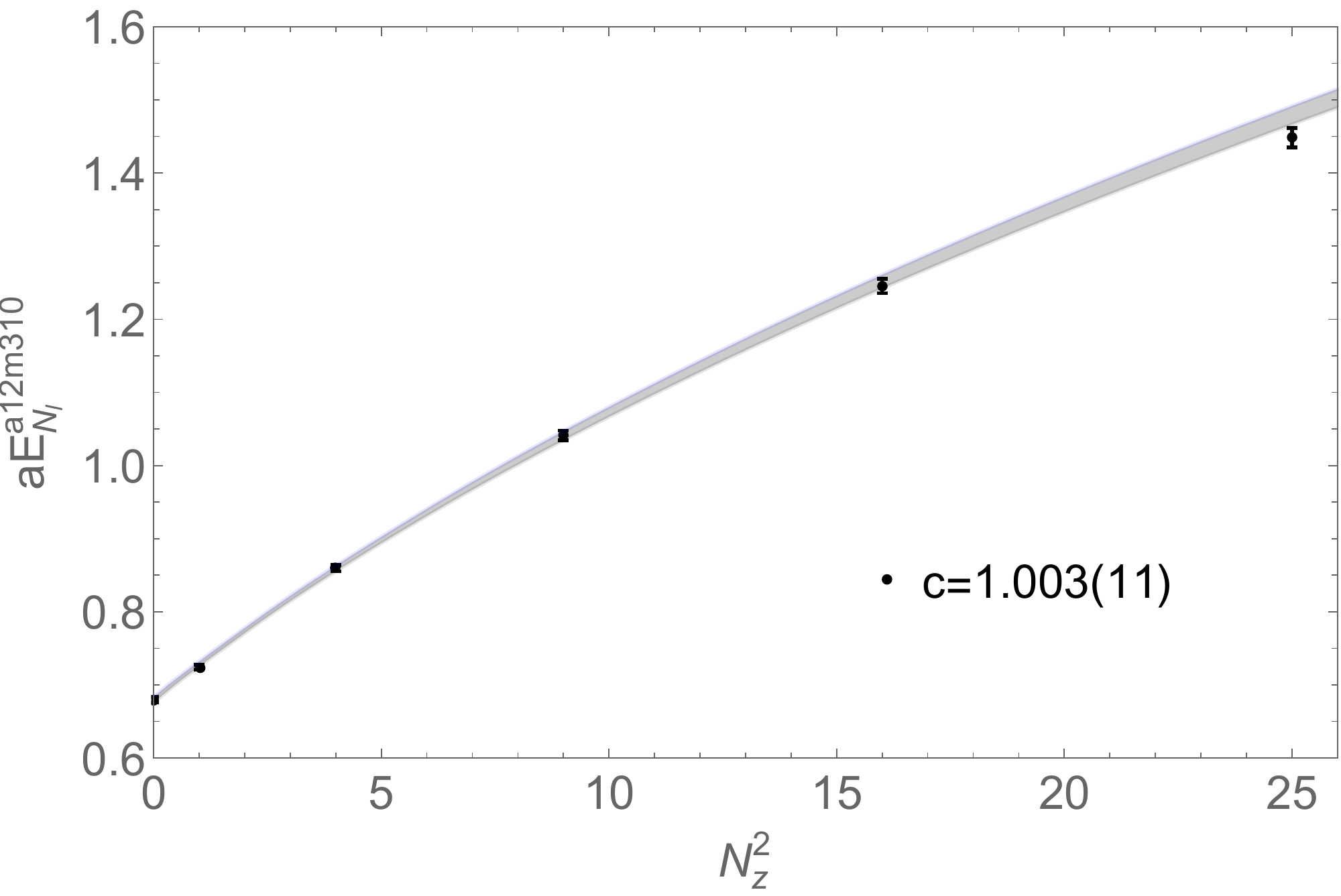}
\caption{Dispersion relations of the nucleon energy from the two-state fits for $M_\pi \approx 690$~MeV (left) and $M_\pi \approx 310$~MeV (right)}
\label{fig:dispersion}
\end{figure*}

The nucleons two-point correlators are then fitted to a two-state ansatz
\begin{align}
C_N^\text{2pt}(P_z,t) &= |A_{N,0}|^2 e^{-E_{N,0}t} + |A_{N,1}|^2 e^{-E_{N,1}t} + ...,
\label{eq:2pt_fit_formula}
\end{align}
where the $ |A_{N,i}|^2$ and $E_{N,i}$ are the ground-state ($i=0$) and first excited state ($i=1$) amplitude and energy, respectively. In this work, we use $N_s$ to denote a nucleon composed of quarks such that $M_\pi \approx 690$~MeV and $N_l$ to denote a nucleon composed of quarks such that $M_\pi \approx 310$~MeV.
Figure~\ref{fig:effectivemass} shows the effective-mass plots for the nucleon two-point functions with $P_z=[0,5]\frac{2\pi}{L}$ for both masses. The bands show the corresponding reconstructed fits using Eq.~\ref{eq:2pt_fit_formula} with fit range $[3, 13]$.
%The reason for this choice of fit range is explained in Fig.~\ref{fig:fitrange-Og} and the corresponding paragraph.
The bands are consistent with the data except where $P_z$ and $t$ are both large. The error of the effective masses at large $P_z$ and $t$ region is too large to fit. However, our reconstructed effective mass bands still match the the data points for the smaller $t$ values even for the largest $P_z=5\times2\pi/L$.
We check the dispersion-relation $E^2=E_0^2+c^2P_z^2$ of the nucleon energy as a function of the momentum, as shown in Fig.~\ref{fig:dispersion}, and the speed of light $c$ for the light quark is consistent with 1 within the statistical errors.

We use the unpolarized gluon operator defined in Ref.~\cite{Balitsky:2019krf},
\begin{align}
{\cal O}_g(z)&\equiv\sum_{i\neq z,t}{\cal O}(F^{ti},F^{ti};z)-\sum_{i,j\neq z,t}{\cal O}(F^{ij},F^{ij};z),
\end{align}
where the operator ${\cal O}(F^{\mu\nu}, F^{\alpha\beta};z) =F^{\mu}{}_{\nu}(z)U(z,0)F^{\alpha}{}_{\beta}(0)$, $z$ is the Wilson link length and the field tensor $F_{\mu\nu}$ is defined as,
\begin{equation}\label{}
\begin{aligned}
 F_{\mu\nu}=\frac{i}{8a^2g_0}(\mathcal{P}_{[\mu,\nu]}+\mathcal{P}_{[\nu,-\mu]}+\mathcal{P}_{[-\mu,-\nu]}+\mathcal{P}_{[-\nu,\mu]}),
\end{aligned}
\end{equation}
where the $a$ is the lattice spacing, $g_0$ is the strong coupling constant, the plaquette $\mathcal{P_{\mu,\nu}}=U_{\mu}(x)U_{\nu}(x+a\hat{\mu})U^{\dag}_{\mu}(x+a\hat{\nu})U^{\dag}_{\nu}(x)$ and $\mathcal{P_{[\mu,\nu]}}=\mathcal{P}_{\mu,\nu}-\mathcal{P}_{\nu,\mu}$.
The operator $O_g$ is chosen because its corresponding matching kernel appears in Ref.~\cite{Balitsky:2019krf}. An alternative operator, $\sum_{i\neq z,t}{\cal O}(F^{ti},F^{zi};z)$, vanishes at $P_z=0$ for kinematic reasons, which would cause additional difficulty in obtaining the distributions from this operator.
We find the bare matrix elements to be consistent with up to 5 HYP-smearing steps, and the signal-to-noise ratios do not improve much with more steps. For the gluon operator used in this paper, we use 4 HYP smearing steps to reduce the statistical uncertainties, as studied in Ref.~\cite{Fan:2018dxu}.

We obtain the three-point gluon correlator by combining the gluon loop with nucleon two-point correlators,
\begin{align}\label{eq:3ptC}
 &C_N^\text{3pt}(z,P_z;t_\text{sep},t) \nonumber \\
 &=\langle 0|\Gamma\int d^3y\, e^{-iyP_z}\chi(\vec y,t_\text{sep}){\cal O}_g(z,t)\chi(\vec 0,0)|0\rangle,
\end{align}
where $t$ is the gluon-operator insertion time, $t_\text{sep}$ is the source-sink time separation, and ${\cal O}_g(z,t)$ is the gluon operator.
The matrix elements of gluon operators can be obtained by fitting the three-point function to its energy-eigenstate expansion,
\begin{equation}
\begin{aligned}
&C_N^\text{3pt}(z,P_z,t,t_\text{sep}) \\
&= |A_{N,0}|^2\langle 0|O_g|0\rangle e^{-E_{N,0}t_\text{sep}} \\
&+ |A_{N,0}||A_{N,1}|\langle 0|O_g|1\rangle e^{-E_{N,1}(t_\text{sep}-t)}e^{-E_{N,0}t} \\
&+ |A_{N,0}||A_{N,1}|\langle 1|O_g|0\rangle e^{-E_{N,0}(t_\text{sep}-t)}e^{-E_{N,1}t} \\
&+ |A_{N,1}|^2\langle 1|O_g|1\rangle e^{-E_{N,1}t_\text{sep}}
+ ...,
\label{eq:3pt_fit_formula}
\end{aligned}
\end{equation}
where the amplitudes and energies, $A_{N,0}$, $A_{N,1}$, $E_{N,0}$ and $E_{N,1}$ are obtained from the two-state fit of the 2-point correlator. $\langle 0|O_g|0\rangle$, $\langle 0|O_g|1\rangle$ ($\langle 1|O_g|0\rangle$), and $\langle 1|O_g|1\rangle$ are the ground state matrix element, ground-excited state matrix element, and excited state matrix element respectively. The ground state matrix element $\langle 0|O_g|0\rangle$ is obtained from either a ``two-sim'' fit, a two-state simultaneous fit on multiple separation times with the $\langle 0|O_g|0\rangle, \langle 0|O_g|1\rangle, \langle 1|O_g|0\rangle$ terms, or a ``two-simRR'' fit, which also includes the $\langle 1|O_g|1\rangle$ term.

\begin{figure*}[htbp]
\centering
	\centering
	\includegraphics[width=0.24\linewidth]{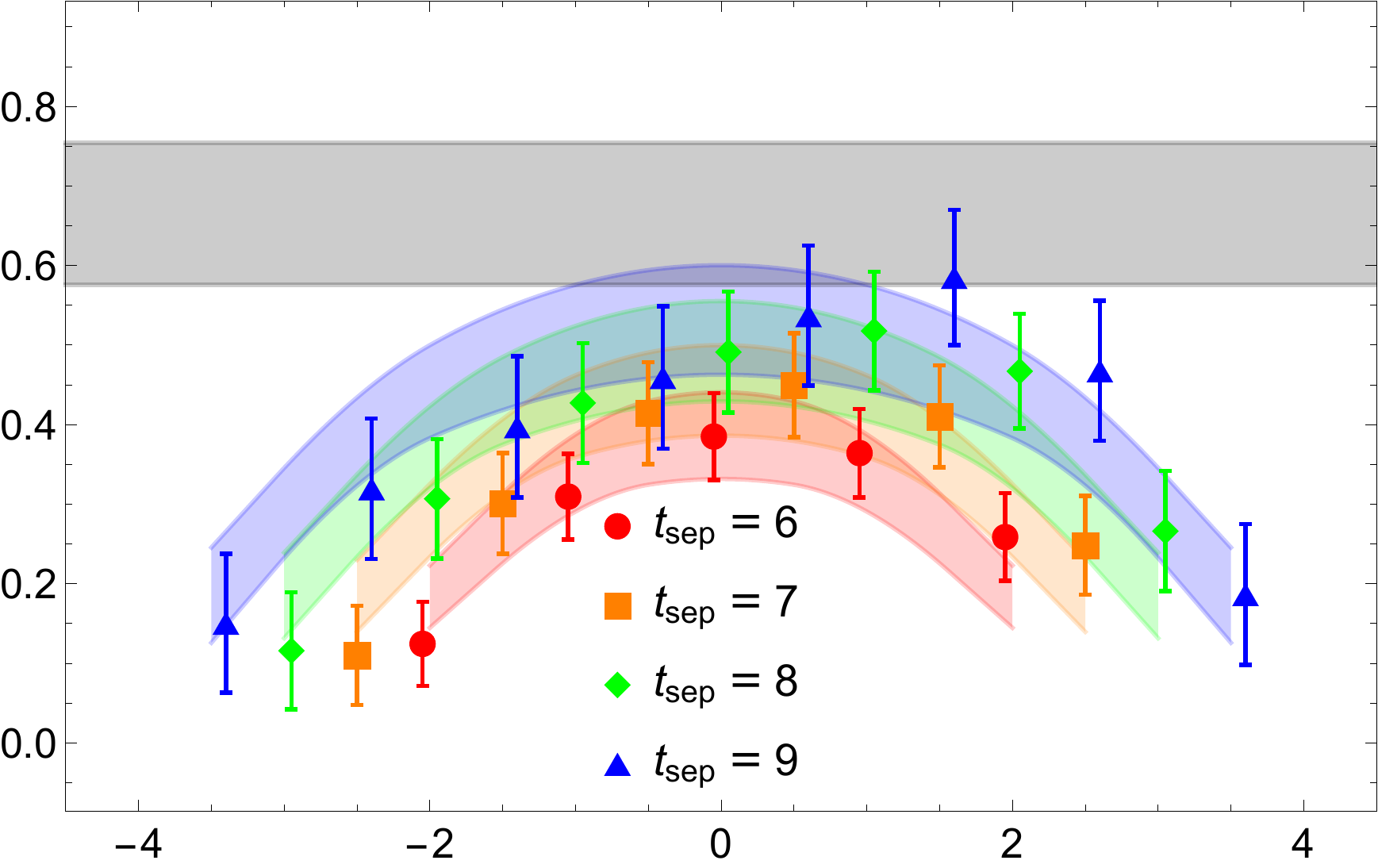}
    \includegraphics[width=0.24\linewidth]{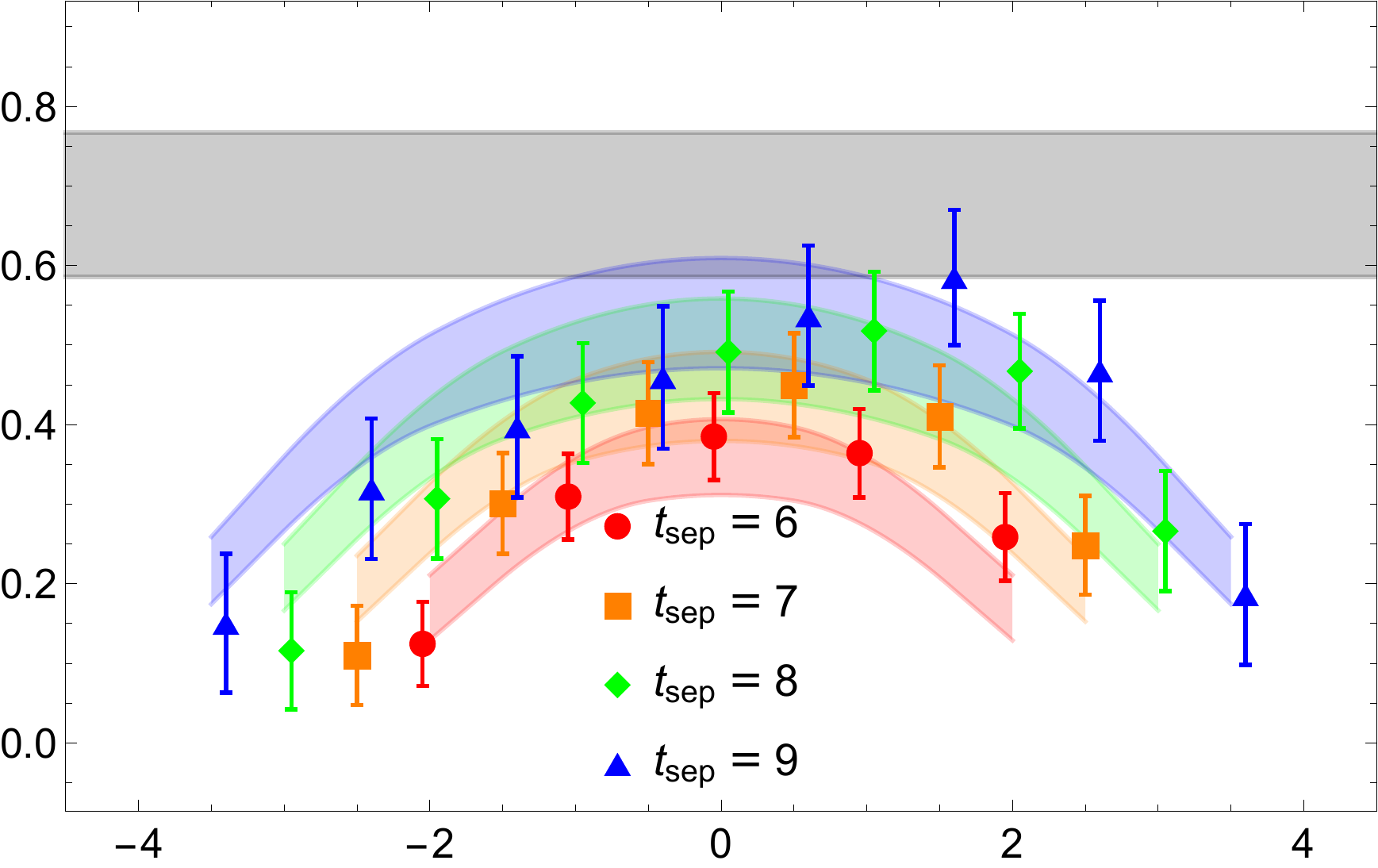}
    \includegraphics[width=0.24\linewidth]{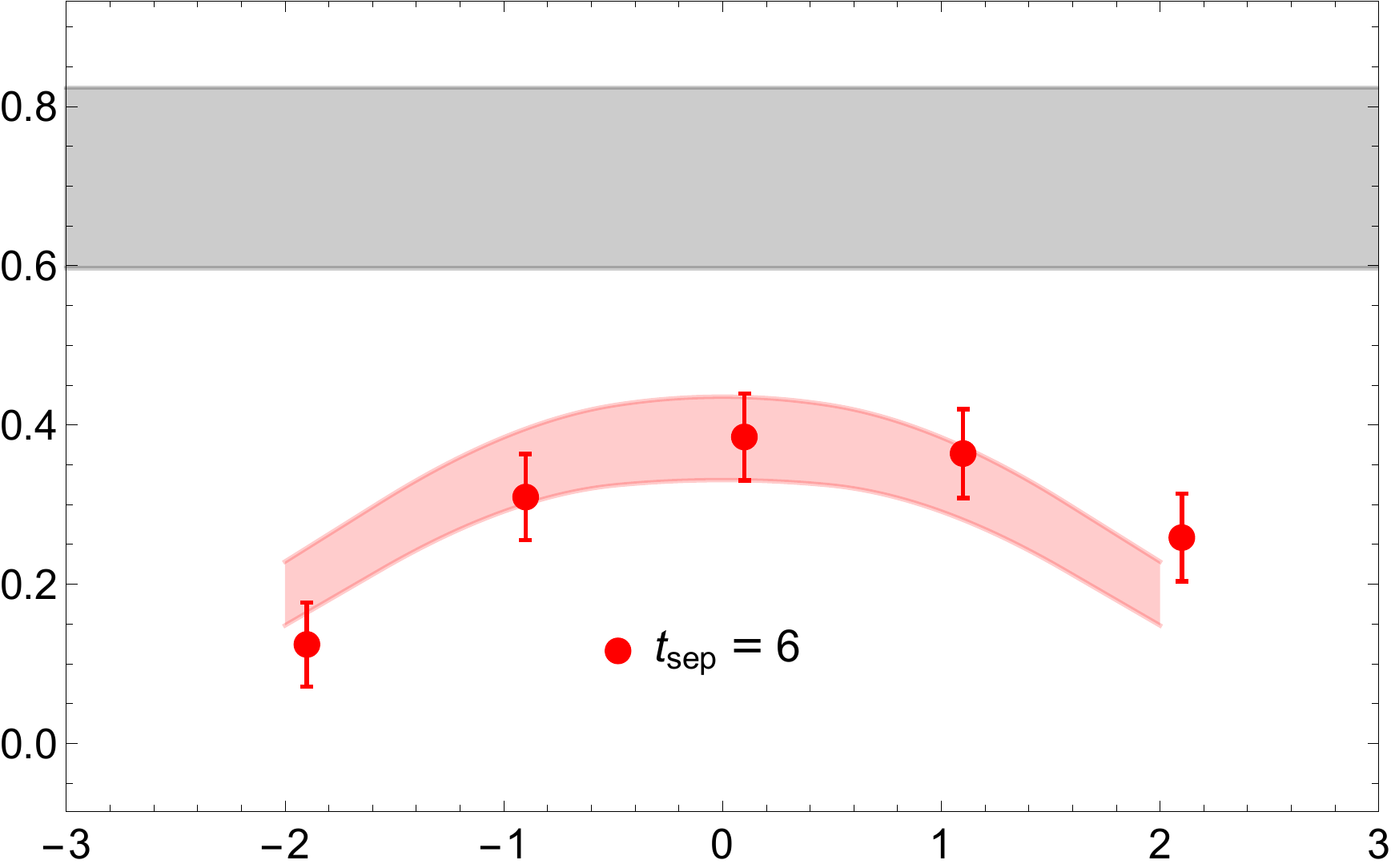}
    \includegraphics[width=0.24\linewidth]{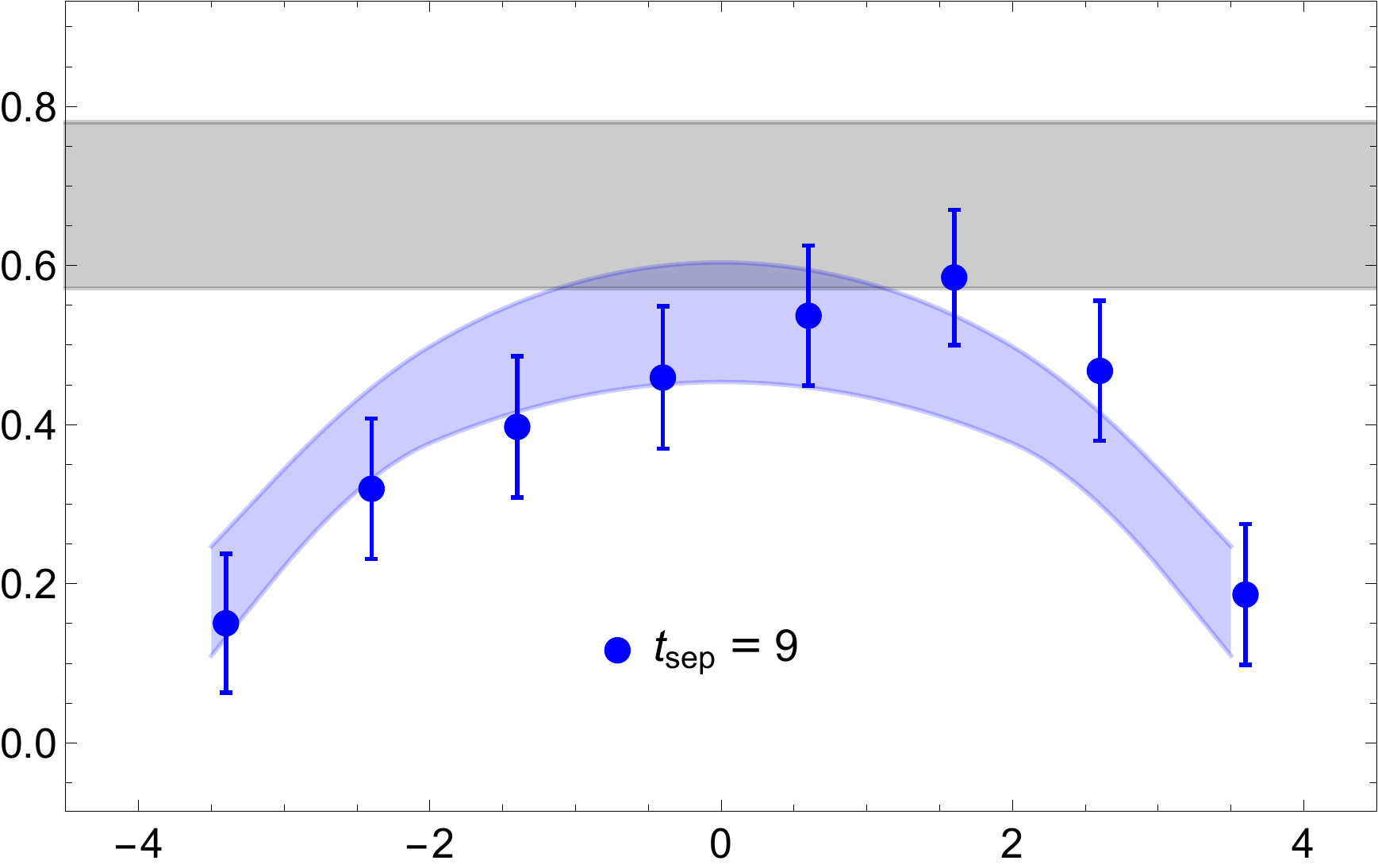}
    \includegraphics[width=0.24\linewidth]{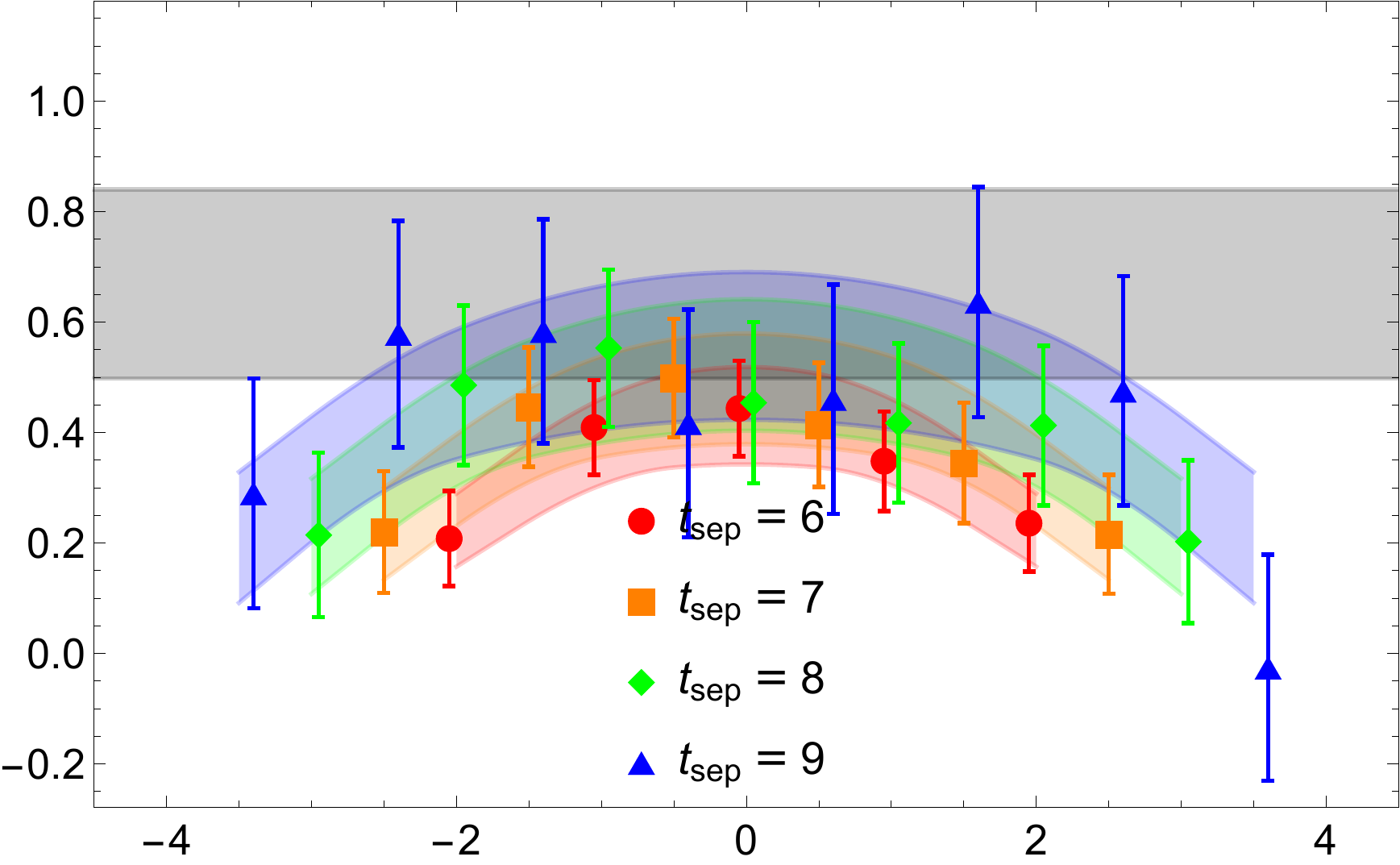}
    \includegraphics[width=0.24\linewidth]{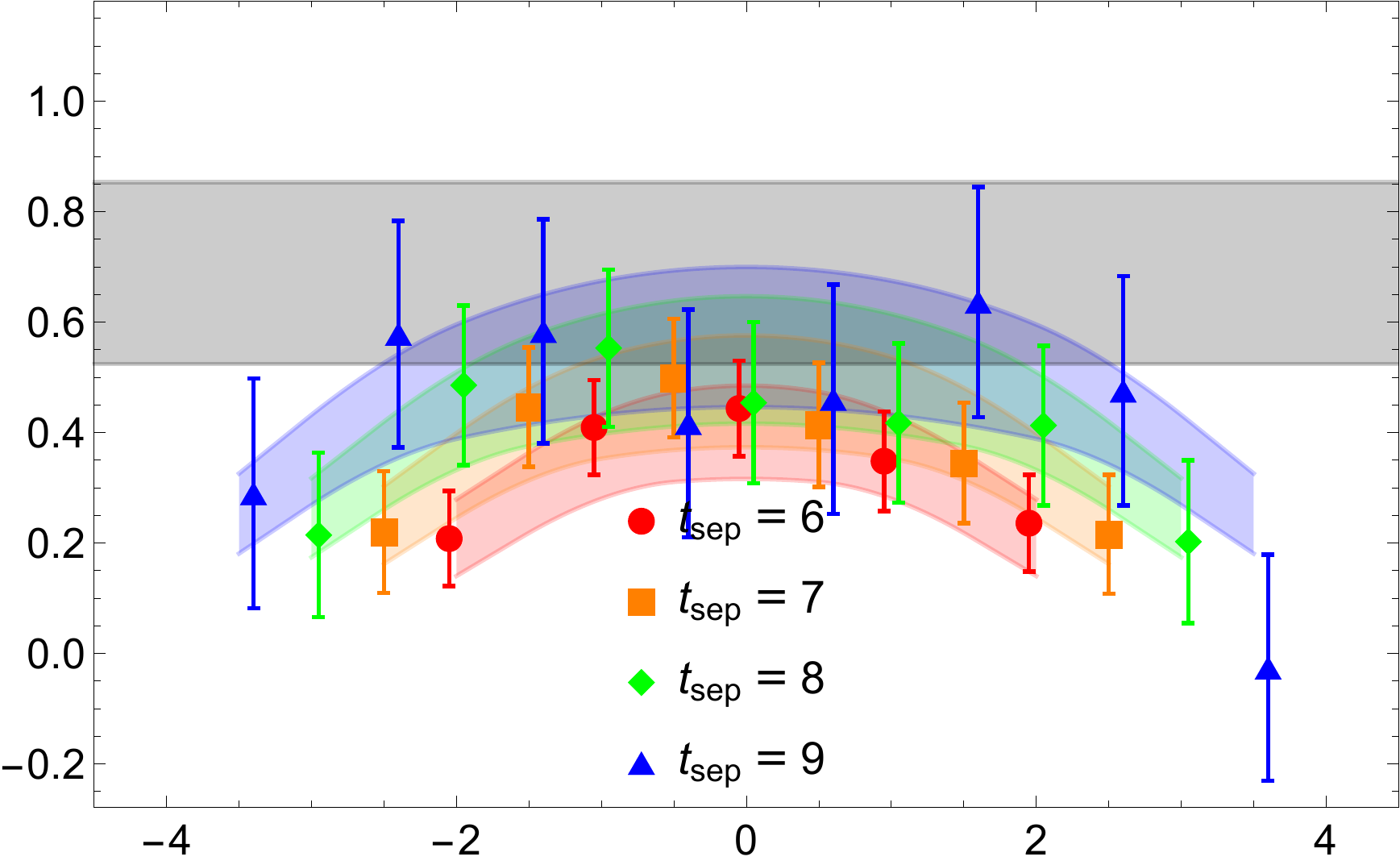}
    \includegraphics[width=0.24\linewidth]{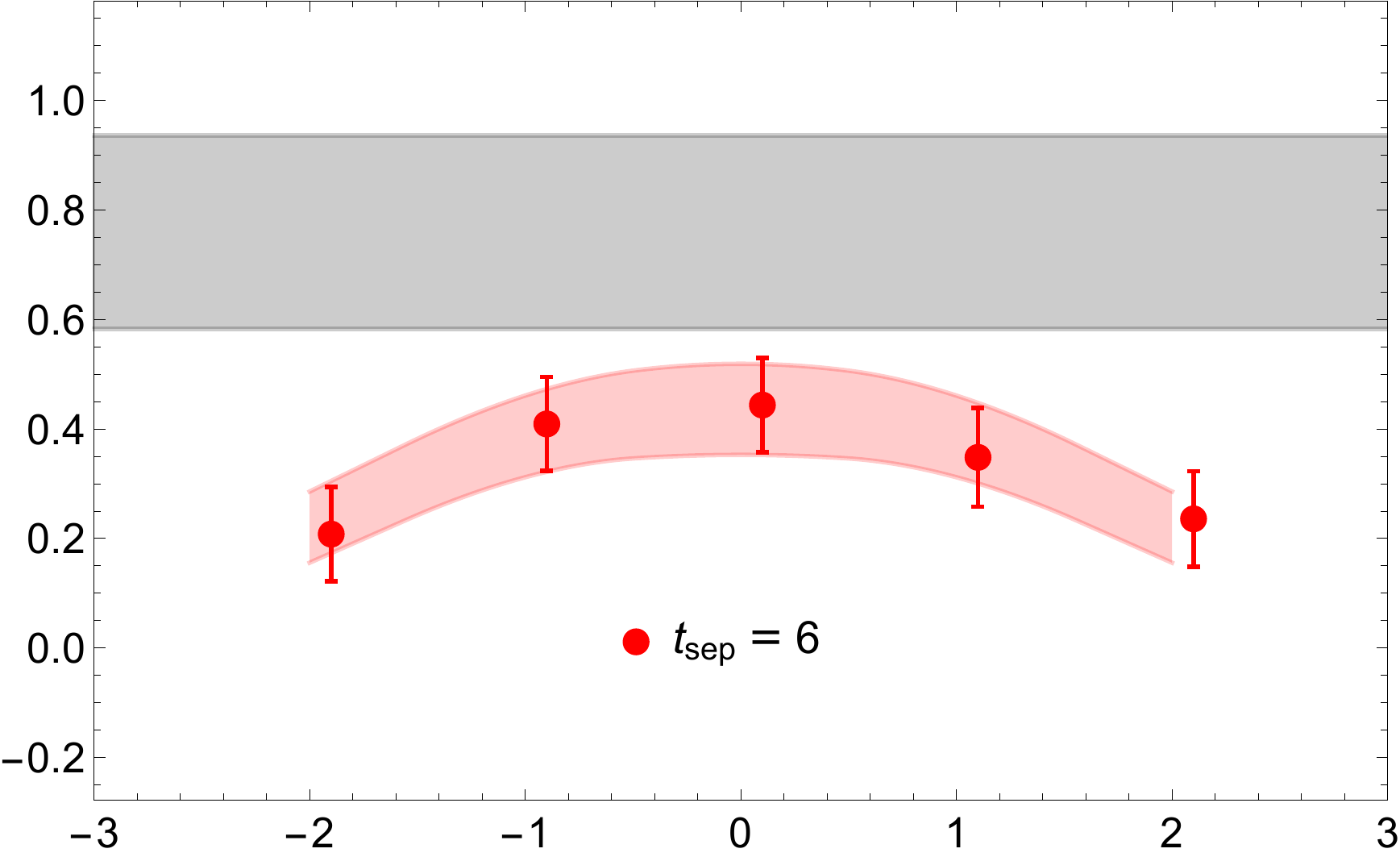}
    \includegraphics[width=0.24\linewidth]{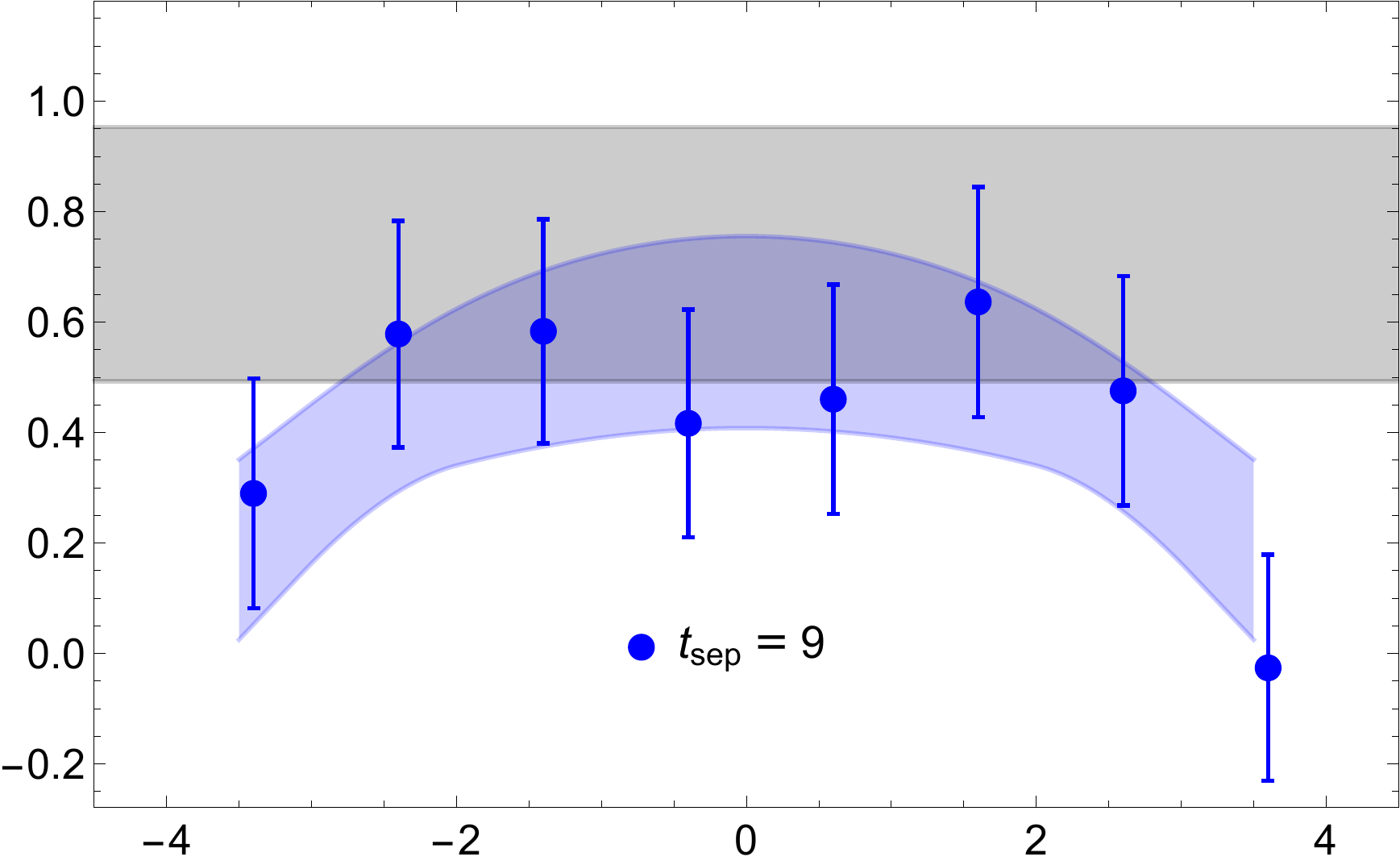}
\caption{The three-point ratio plots for $M_\pi \approx 690$~MeV (top row) and $M_\pi \approx 310$~MeV (bottom row)% two rows)
nucleons $z=1$ as functions of $t-t_\text{sep}/2$, as defined in Eq.~\ref{eq:ratio_def}.
The results for nucleon momentum %$P_z\in \{2,2,5\}\times 2\pi/L$
$P_z = 2\times 2\pi/L$ are shown.
The gray bands in each panel indicate the extracted ground-state matrix elements of the operator $O_g$.
In each column, the plots show the fitted ratio and the extracted ground-state matrix elements from two-simRR and two-sim fits with all 4 source-sink separations, and the two-state fits using only the smallest and largest $t_\text{sep}$ from left to right, respectively. The second column, which are the two-sim extracted ground-state matrix elements, are used in the subsequent analysis.
The ground-state matrix elements extracted are stable and consistent among different fitting methods and three-point data input used.
}
\label{fig:fitcomp-Og}
\end{figure*}

Figure~\ref{fig:fitcomp-Og} shows example correlator plots from the ratio
\begin{align}
R_N(P_z,t,t_\text{sep})=\frac{C_N^\text{3pt}(P_z,t,t_\text{sep})}{C_N^\text{2pt}(P_z,t_\text{sep})}.
\label{eq:ratio_def}
\end{align}
as a function of the $t-t_\text{sep}/2$ for multiple source-sink separations
for at %$P_z\in\{2,5\}\times2\pi/L$
$P_z=2\times 2\pi/L$
and $t_\text{sep}=\{6,7,8,9\}\times a$.
The reconstructed ratio plot, using the fitted parameters obtained from Eqs.~\eqref{eq:3pt_fit_formula} and \eqref{eq:2pt_fit_formula} are plotted for each $t_\text{sep}$, and the gray band indicates the reconstructed ground-state matrix elements $\langle 0|{\cal O}_g|0\rangle$.
The left-two plots in Fig.~\ref{fig:fitcomp-Og} show the two-simRR fits and two-sim fits using the $t_\text{sep}=\{6,7,8,9\} a$, while the remaining two plots show individual two-state fits to the smallest and largest source-sink separations ($t_\text{sep}=\{6,9\}a$).
The plots of pion mass $M_\pi \approx 690$~MeV and $M_\pi \approx 310$~MeV are shown in the first row and second row respectively. The reconstructed ground state matrix elements (gray bands) for $O_{g}$ are consistent for the fits with individual $t_\text{sep}=\{6,9\}$, the two-sim fit results and the two-simRR fit within one sigma error.
Therefore, the two-sim fits describe data from $t_\text{sep}=\{6,7,8,9\}$ well for operator $O_{g}$. Thus, we use the two-sim fits to extract the ground-state matrix element $\langle 0|O_g|0\rangle$ of different $z$, $P_z$ for the rest of this paper.

%Figure~\ref{fig:ratio-Og} shows the ratio of three-point to two-point correlators and the reconstructed band of a two-state fit of operator $O_{g}$ for the $M_\pi \approx 690$~MeV and $M_\pi \approx 310$~MeV nucleons on the a12m310 ensemble with $P_z \in \{2,5\}\times2\pi/L$, $z=5a$. The fit results are stable even for the largest momentum $P_z=5\times2\pi/L$ with the largest error. The fitted bare matrix element $\langle 0|{\cal O}_g|0\rangle$ is shown in Tab.~\ref{tab:correlator-fit}. We present examples of the extracted ground state matrix elements at both large and small momentum $P_z$ and Wilson link length $z$ from the two-state fits.

%\begin{table}[!htbp]
%\centering
%\begin{tabular}{|c|c|c|c|}
%\hline
%  $M_\pi$ (MeV)  & $z$ ($a$) & $N_z$  &   $\langle 0|{\cal O}_g|0\rangle$ \\
%  \hline
%  690 & 1 & 2 & $0.68(10)$   \\
%   & 5 & 2 & $0.191(54)$   \\
%   & 1 & 5 & $1.16(35)$   \\
%   & 5 & 5 & $0.22(13)$  \\
%  \hline
%  310 & 1 & 2 & $0.72(23)$   \\
%   & 5 & 2 & $0.209(98)$   \\
%   & 1 & 5 & $1.09(91)$  \\
%   & 5 & 5 & $-0.15(33)$  \\
%\hline
%\end{tabular}
%\caption{The $M_\pi \approx 690$~MeV and $M_\pi \approx 310$~MeV nucleons three-point correlators fitted bare matrix element $\langle 0|{\cal O}_g|0\rangle$ at $z\in\{1,5\}\times a, P_z = N_z\times2\pi/L \in \{2,5\}\times2\pi/L$.}
%\label{tab:correlator-fit}
%\end{table}

\begin{figure*}[htbp]
\centering
	\centering
	\includegraphics[width=0.32\linewidth]{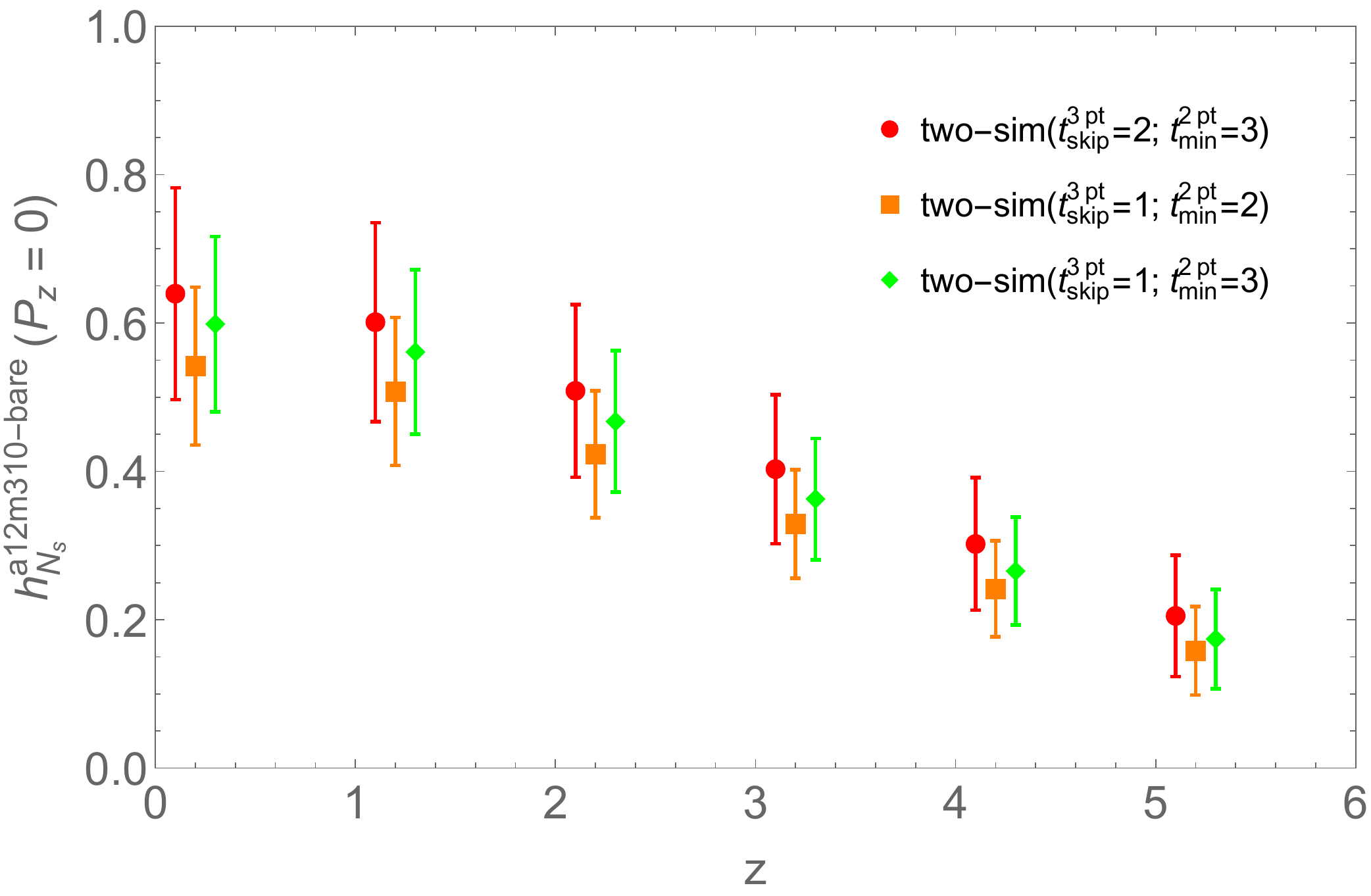}
	\includegraphics[width=0.32\linewidth]{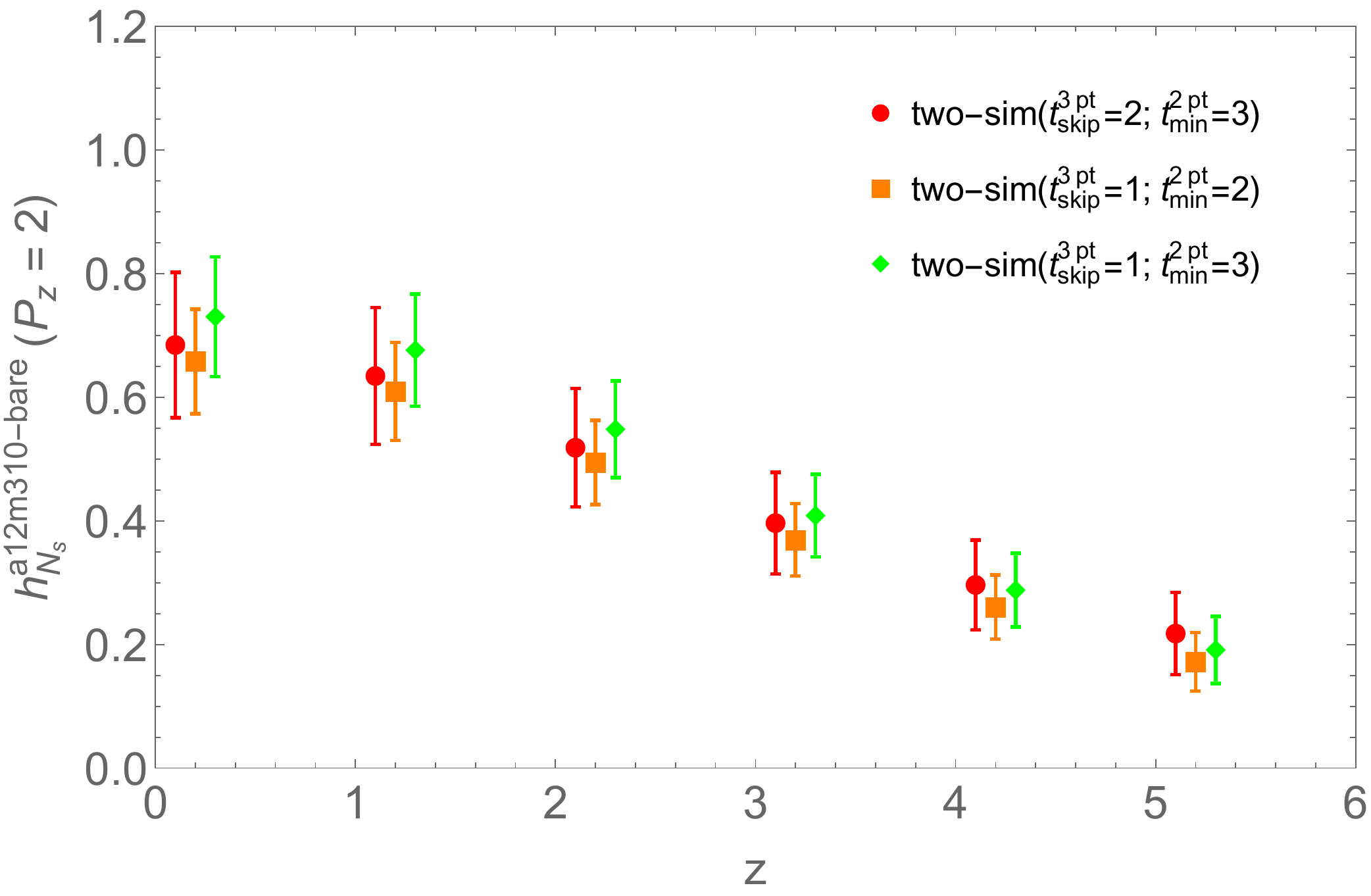}
	\includegraphics[width=0.32\linewidth]{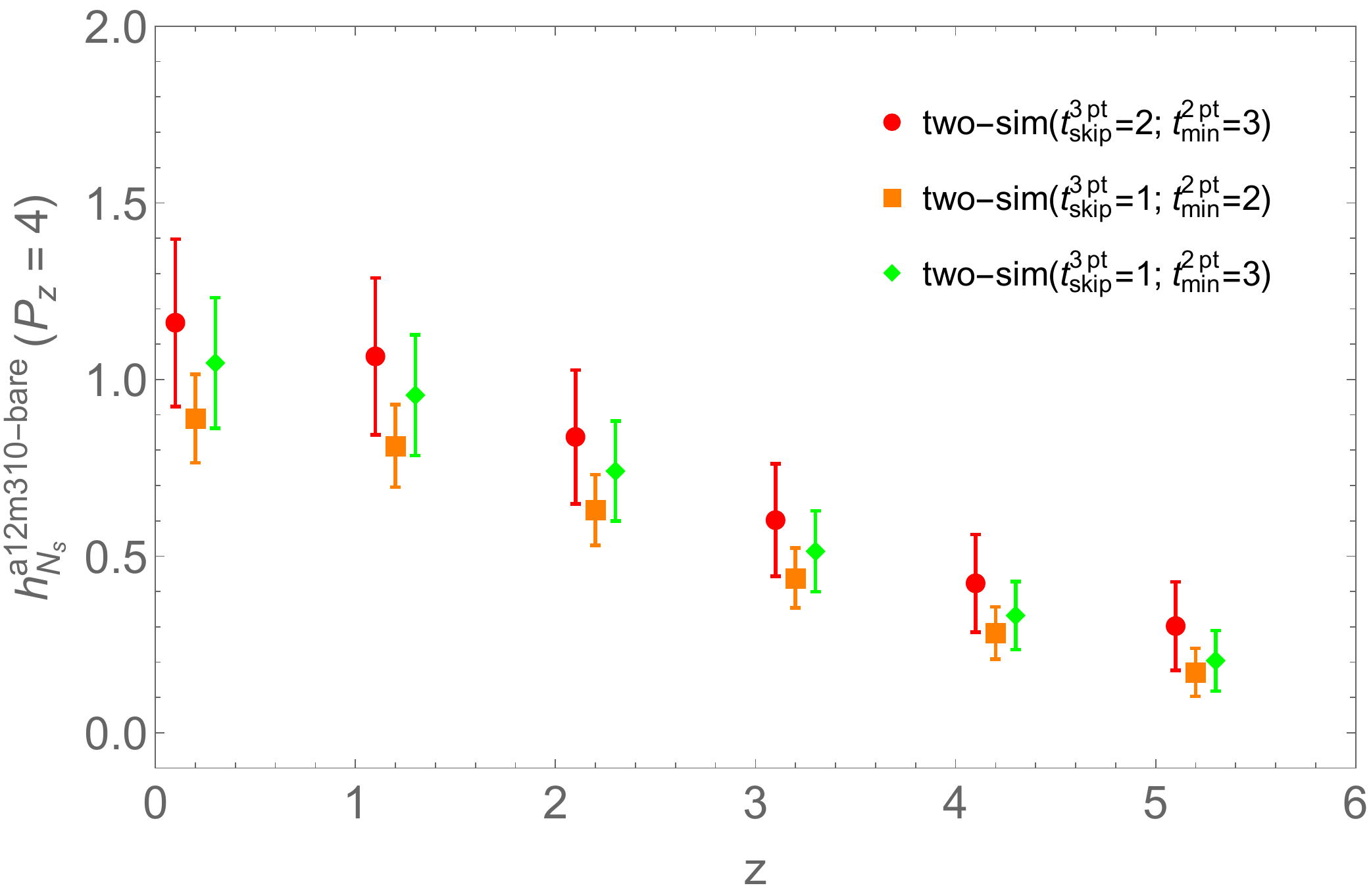}
	\includegraphics[width=0.32\linewidth]{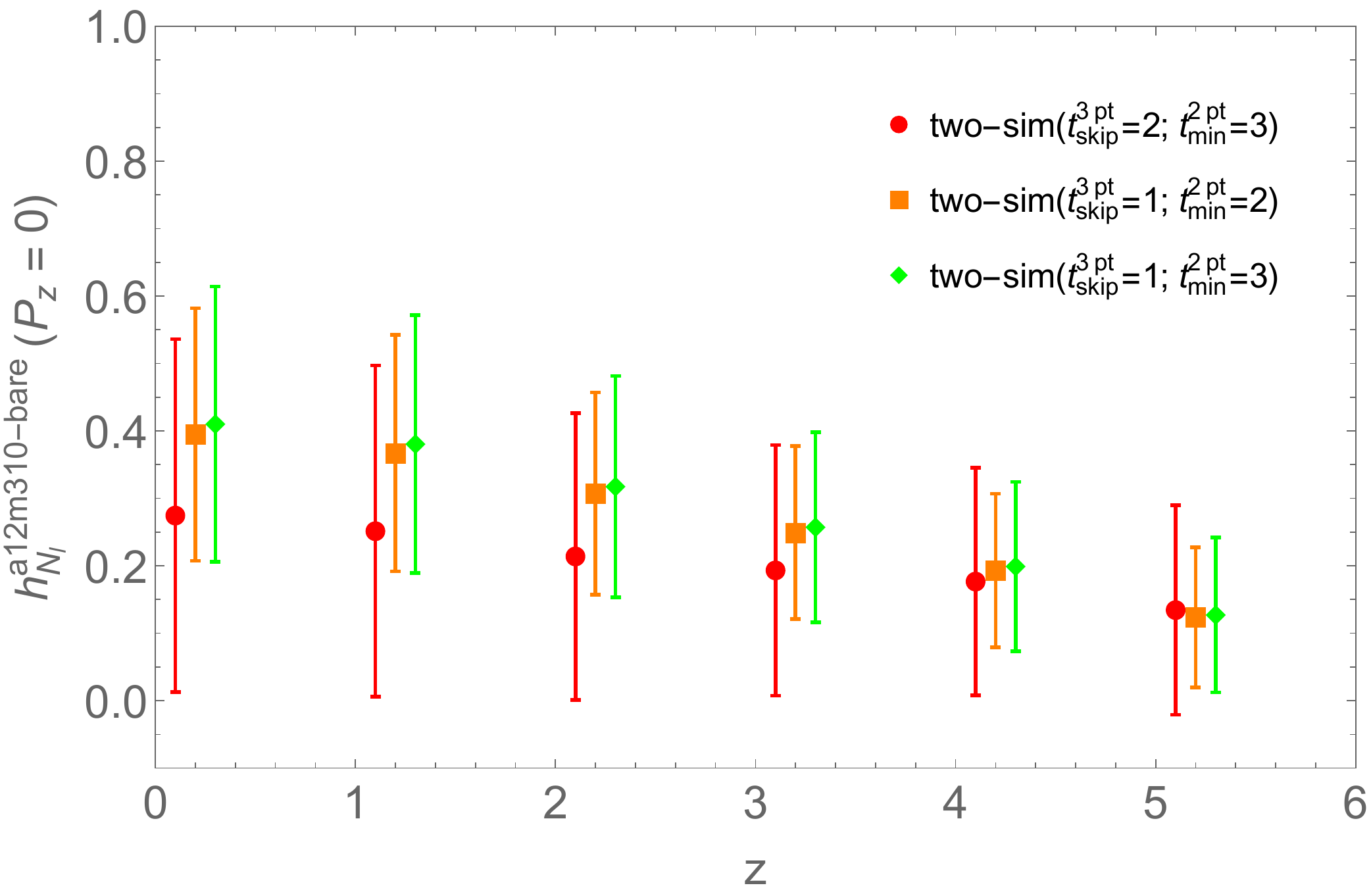}
	\includegraphics[width=0.32\linewidth]{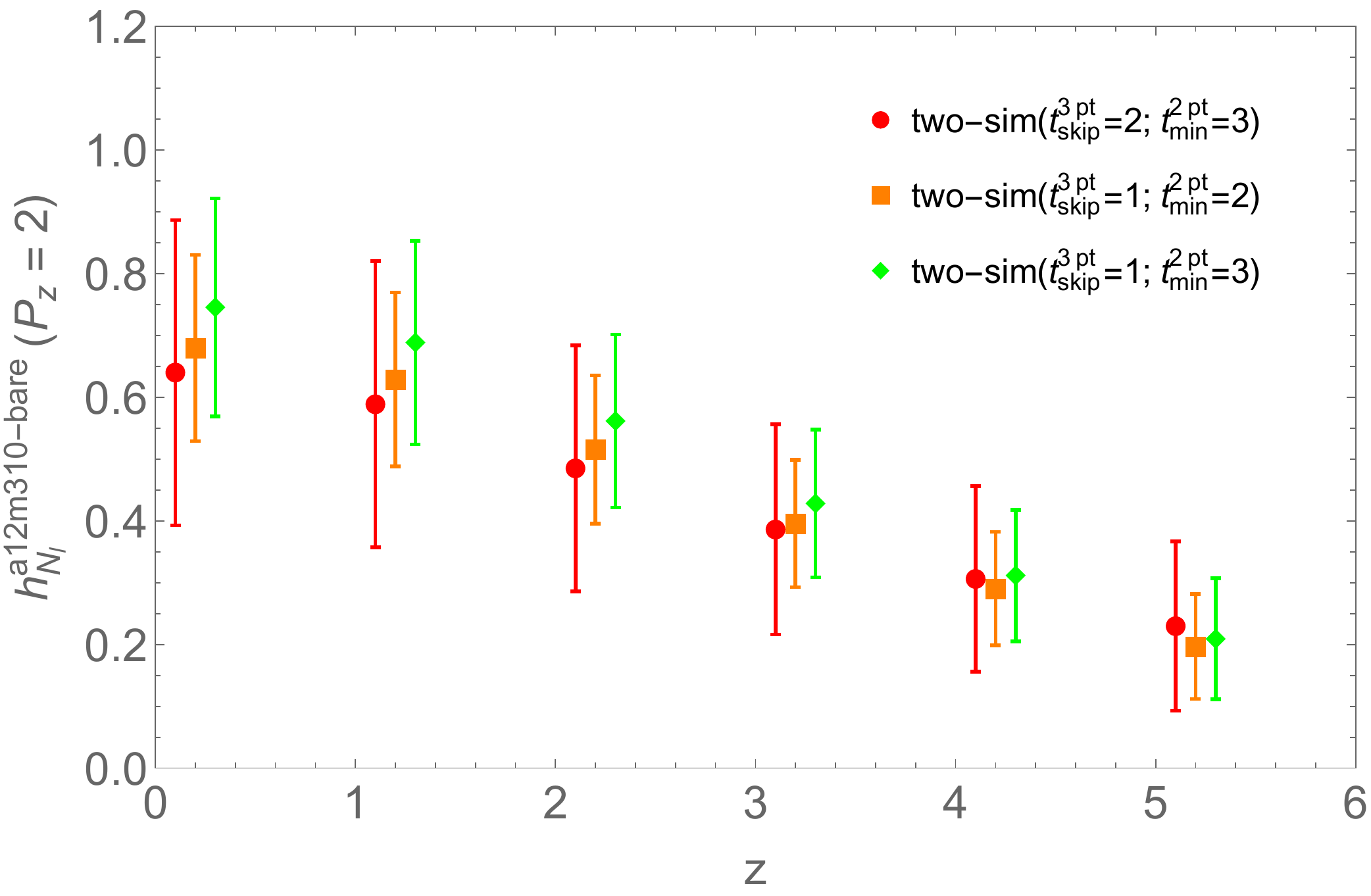}
	\includegraphics[width=0.32\linewidth]{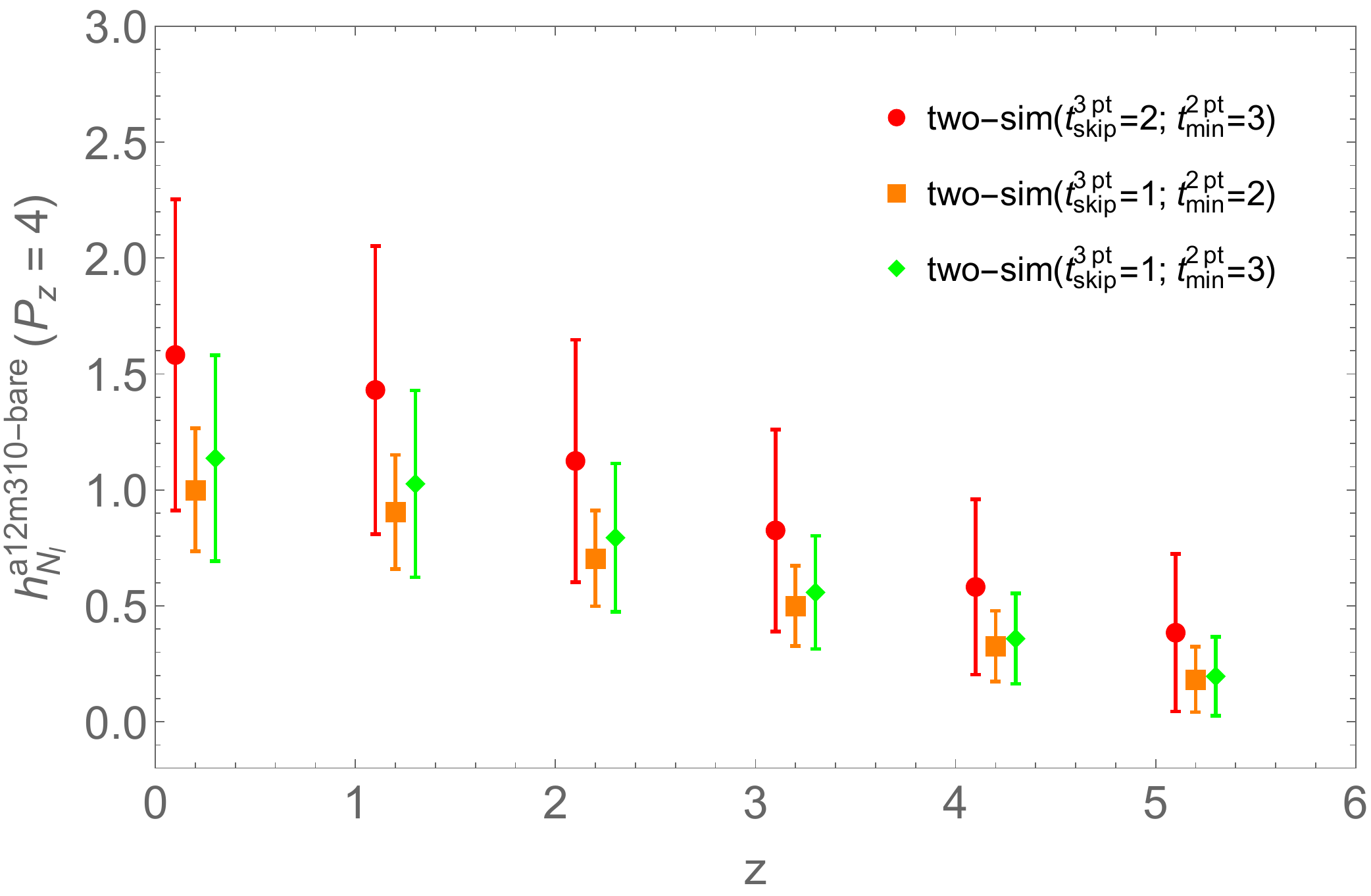}
\caption{The fitted bare ground-state matrix elements without normalization by kinematic factors as functions of $z$ obtained from the two-sim fit using different two- and three-point fit ranges for nucleon momentum $P_z \in \{0,2,4\}\times2\pi/L$ from left to right, respectively, for $M_\pi \approx 690$~MeV (first row) and $M_\pi \approx 310$~MeV (second row) nucleons. The green points, which represent the fit-range choice $t_\text{skip}^\text{3pt}=1$, $t_\text{min}^\text{2pt}=3$ are used in the following analysis, because the errors of the matrix elements of this fit range are relatively smaller than the error of the red points. The orange points, which represent the fit-range choice $t_\text{skip}^\text{3pt}=1$, $t_\text{min}^\text{2pt}=2$, are not used because the $\chi^2/\text{dof}$ of the 2-point correlator fits with $t_\text{min}^\text{2pt}=2$ are much larger than $t_\text{min}^\text{2pt}=3$ cases.
}
\label{fig:fitrange-Og}
\end{figure*}

Our extracted bare ground-state matrix elements are stable across various fit ranges. Figure~\ref{fig:fitrange-Og} shows example results from $M_\pi \approx 690$~MeV and $M_\pi \approx 310$~MeV nucleons with nucleon momentum $P_z \in [1,5]\times2\pi/L$ as the fit ranges for two- and three-point varies.
In this case, the two-point correlator fit ranges are $[t_\text{min}, 13]$ and the three-point correlators fit ranges are $[t_\text{skip}, t_\text{sep}-t_\text{skip}]$.
All the matrix elements from different fit ranges are consistent with each other in one-sigma error.
The fit range choice $t_\text{skip}^\text{3pt}=1$, $t_\text{min}^\text{2pt}=2$ are not used, because the $\chi^2/\text{dof}$ of the 2-point correlator fits with $t_\text{min}^\text{2pt}=2$ are much larger than the $t_\text{min}^\text{2pt}=3$ cases.
For the rest of this paper, we use the fitted matrix elements obtained from the fit-range choice $t_\text{skip}^\text{3pt}=1$, $t_\text{min}^\text{2pt}=3$.
The extracted bare matrix elements are fitted for $P_z \in [0,5]\times2\pi/L$ and $z \in [0,5]\times a$ to obtain the Ioffe-time distributions in pseudo-PDF calculation.

%\begin{figure*}[htbp]
%\centering
%	\centering
%	\includegraphics[width=0.23\linewidth]{figs/Ratio_c898_O0IB_dir-average_p2_z5_Re_a12m690.pdf}
%	\includegraphics[width=0.23\linewidth]{figs/Ratio_c898_O0IB_dir-average_p5_z5_Re_a12m690.pdf}
%	\includegraphics[width=0.23\linewidth]{figs/Ratio_c898_O0IB_dir-average_p2_z5_Re_a12m310.pdf}
%	\includegraphics[width=0.23\linewidth]{figs/Ratio_c898_O0IB_dir-average_p5_z5_Re_a12m310.pdf}
%\caption{The ratio of the $O_{g}$ three-point correlators to the two-point correlators and the band of the two-sim fit on the a12m310 ensemble for $M_\pi \approx 690$~MeV (first two plots) and $M_\pi \approx 310$~MeV (last two plots) nucleons at $z=5a$, $P_z\in \{2,2,5,5\}\times 2\pi/L$ (from left to right).}
%\label{fig:ratio-Og}
%\end{figure*}

%%%%%%%%%%%%%%%%%%%%%%%%%%%%%%%%%%%%%%%%%%%%%%%%%%%%%%%%%%%%%%%%%%%%%%%%%%%%%%%%
\section{Results and Discussions}
\label{sec:results}

In the previous section, we obtained the gluon ground-state bare matrix element $\langle 0|{\cal O}_g(z)|0\rangle$ at different $z$ and $P_z$. The Ioffe-time distribution (ITD) $\mathcal{M}(\nu,z^2)$ is
\begin{align}
\mathcal{M}(\nu,z^2) = \langle 0|{\cal O}_g(z)|0\rangle,
\label{eq:ME_unpol}
\end{align}
where Ioffe time $\nu=zP_z$. We construct the reduced ITD, where we take the ratio of the ITD with its value at $\nu=0$, to eliminate ultraviolet divergences.
We then further normalize the ratio by the reduced ITD at $z^2=0$
to cancel out the kinematic factors and improve the signal-to-noise ratios. The resulting double ratio~\cite{Orginos:2017kos} is
\begin{align}
\mathscr{M}(\nu,z^2)=\frac{\mathcal{M}(\nu,z^2)/\mathcal{M}(\nu,0)}{\mathcal{M}(0,z^2)/\mathcal{M}(0,0)}.
\label{eq:RITD}
\end{align}
Dividing up the matrix elements by their corresponding boost momentum at $z=0$ also has the advantages of reducing the statistical and lattice systematic errors, and has been done since the first Bjorken-$x$--dependent PDF calculation~\cite{Lin:2013yra} in 2013. One of the reasons we use operator $O_g$ is that unlike other nonperturbatively renormalizable operators, it gives nonzero results at $P_z=0$. This makes it a better choice to form the reduced pseudo-ITD by taking a double ratio, as discussed in Ref.~\cite{Balitsky:2019krf}.
The reduced-ITD double ratios used here have no additional explicit normalization~\cite{Orginos:2017kos}, and one can apply the pseudo-PDF matching condition~\cite{Balitsky:2019krf} to obtain the unpolarized gluon PDF,
\begin{align}
\mathscr{M}(\nu,z^2)=\int_0^1 dx \frac{xg(x,\mu^2)}{\langle x_g \rangle_{\mu^2}}R(x\nu,z^2\mu^2),
\end{align}
where $\mu$ is the renormalization scale in $\overline{\text{MS}}$ scheme and $\langle x_g \rangle_{\mu^2}=\int_0^1 dx \ x f_g(x,\mu)$ is the gluon momentum fraction of the nucleon.
The matching kernel, $R(y=x\nu,z^2\mu^2)$, is composed of two terms to deal with the effects of evolution and scheme conversion~\cite{Radyushkin:2017cyf},
\begin{align}
&R(y,z^2\mu^2)=R_1(y,z^2\mu^2)+R_2(y), \label{matching-kernel}\\
&R_1(y,z^2\mu^2)=-\frac{\alpha_s(\mu)}{2\pi}N_c\ln\left(z^2\mu^2\frac{e^{2\gamma_E+1}}{4}\right)R_B(y), \label{matching-evolve}\\
&R_2(y)=\cos y-\frac{\alpha_s(\mu)}{2\pi}N_c\left( 2R_B(y)+R_L(y)+R_C(y)\right),
\label{matching-scheme}
\end{align}
where $R_1(y,z^2\mu^2)$ is the term related to evolution, $R_2(y,z^2\mu^2)$ is the term related to scheme conversion, $\alpha_s$ is the strong coupling at scale $\mu$, $N_c=3$ is the number of colors, $\gamma_E=0.5772$ is Euler-Mascheroni constant, and $R_B(y)$, $R_L(y)$ and $R_C(y)$ are defined in Eqs.~7.21--23 in Ref.~\cite{Balitsky:2019krf}. The $z$ in $R_1(y,z^2\mu^2)$ is chosen to be $e^{-\gamma_e-1/2}/\mu$ so that the log term vanishes, suppressing the residuals that contain higher orders of the log term, as discussed in Ref.~\cite{Radyushkin:2018cvn}.

\begin{figure*}[htbp]
\centering
	\centering
	\includegraphics[width=0.4\linewidth]{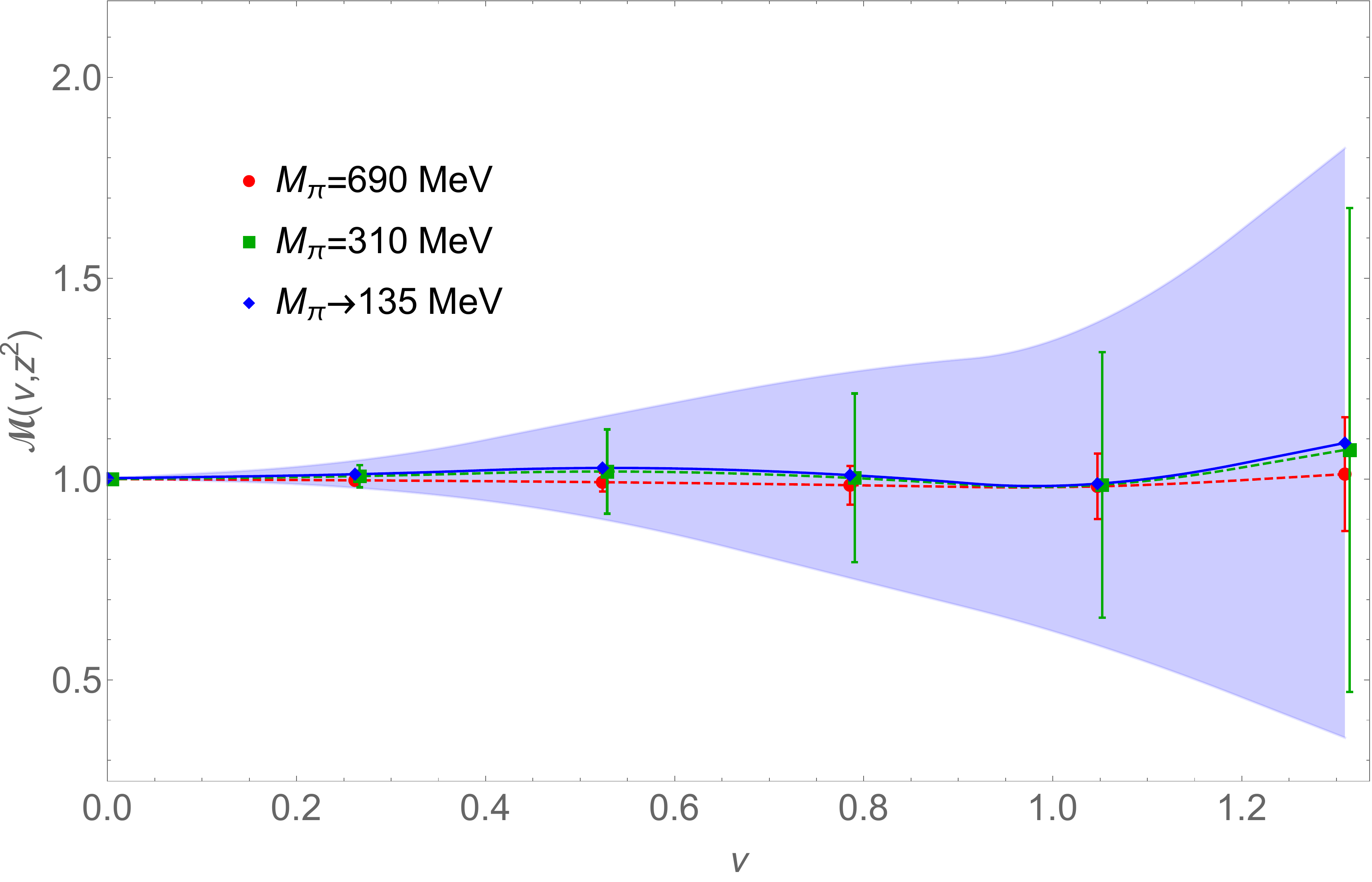}
	\includegraphics[width=0.4\linewidth]{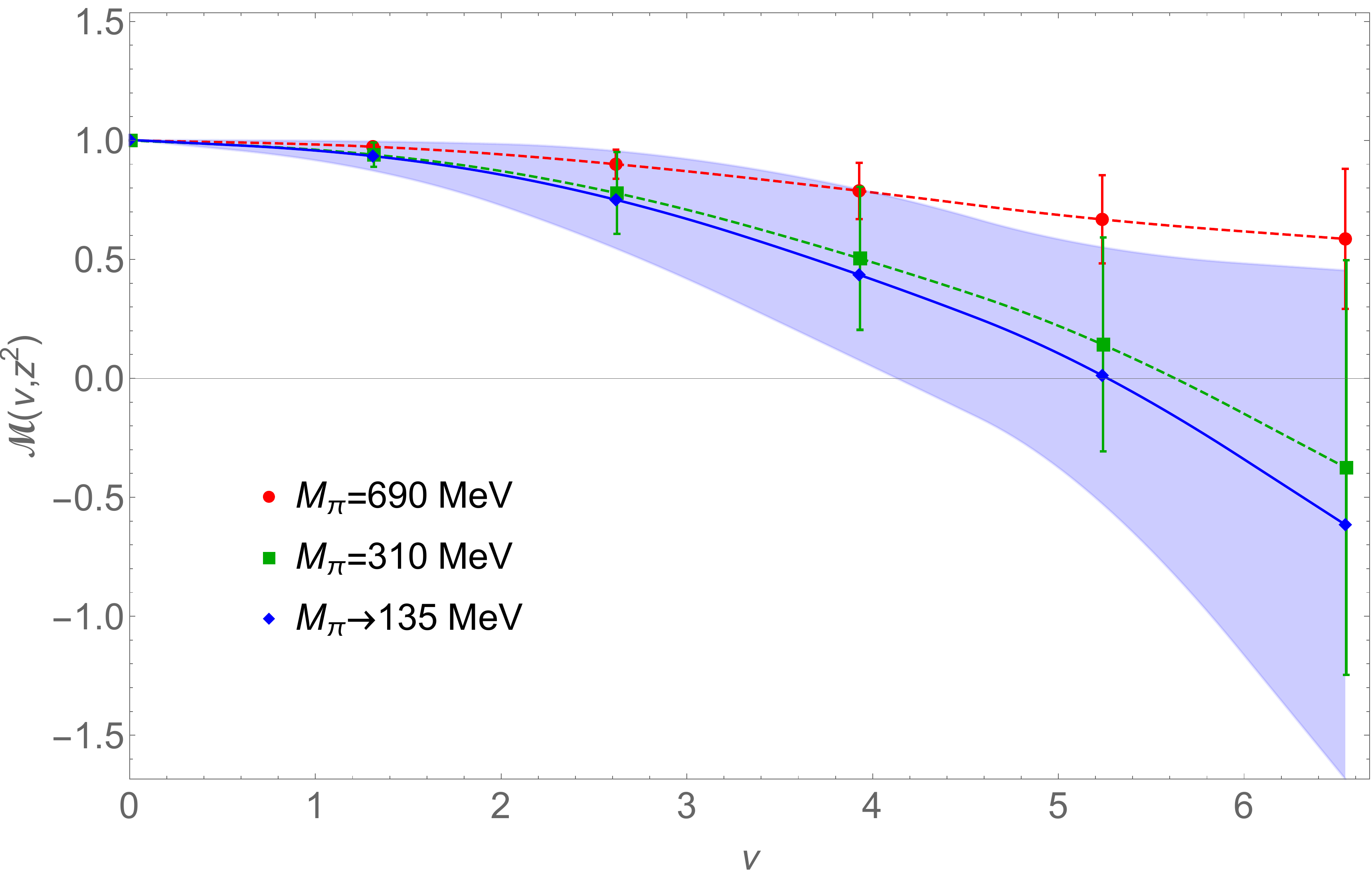}
\caption{The reduced ITDs $\mathscr{M}(\nu, z^2)$ as functions of $\nu$ and their extrapolation to the physical pion mass at $P_z=1\times 2\pi/L$ (left) and $P_z=5\times 2\pi/L$ (right). The blue bands represent the fitted results of the reduced ITDs at the physical pion mass $M_\pi = 135$~MeV.
}
\label{fig:extrapolation}
\end{figure*}

The lightcone PDF at physical pion mass is obtained from the reduced ITDs by the following procedure. First, we extrapolate the reduced ITDs to physical pion mass. Second, we evolve the reduced ITDs. Finally, we assume a functional form for the unpolarized gluon PDF and use the matching kernel to match it to the evolved ITDs to fit with the lattice simulation data.
In order to determine the gluon PDF at physical pion mass, we extrapolate our reduced ITD results at $M_\pi =690$ and 310~MeV to $M_\pi^\text{phys}=135$~MeV  using the following simple naive ansatz:
\begin{align}
\mathscr{M}(\nu,z^2,M_{\pi})&=\mathscr{M}(\nu,z^2,M_\pi^\text{phys}) \nonumber \\
&+K(\nu,z^2)(M_{\pi}^2-(M_\pi^\text{phys})^2),
\label{extrapolation}
\end{align}
where $M_\pi^\text{phys}$ is the physical pion mass. We fit the reduced ITDs for each jackknife sample at each $P_z$ and $z$ value. The slope $K$ is about $-0.05\text{ GeV}^{-2}$ in our fit. Then, the jackknife samples of the reduced ITDs at physical pion mass are reconstructed from the fit parameters from each jackknife sample fit. Figure~\ref{fig:extrapolation} shows the extrapolation results for the reduced ITDs at $P_z\in\{1,5\}\times 2\pi/L$.

The evolved ITD $G$ is obtained by using the evolution term $R_1(y,z^2\mu^2)$ in Eq.~\ref{matching-evolve},
\begin{align}
G(\nu,z^2,\mu,M_{\pi})&=\mathscr{M}(\nu,z^2,M_{\pi})\nonumber\\
&+\int^1_0 \text{d}u R_1(u,z^2\mu^2)\mathscr{M}(u\nu,z^2,M_{\pi}).
\label{evolution}
\end{align}
To obtain the evolved ITD, we interpolate the reduced ITD $\mathscr{M}(\nu,z^2)$ to be a continuous function of $\nu$, using
%There are different ways to achieve this goal, like the the polynomial expansion fit in a pseudo PDF calculation paper of unpolarized nucleon PDF~\cite{Joo:2019jct} and
``$z$-expansion''\footnote{Note that the $z$ in the ``$z$-expansion'' is not related to the Wilson link length $z$ we use elsewhere.} fit~\cite{Boyd:1994tt,Bourrely:2008za} (also adopted by past pseudo-PDF calculations~\cite{Joo:2019bzr})
%paper of pion valence PDF ~\cite{Joo:2019bzr}. The following ``$z$-expansion'' form is used in our calculation,
%
\begin{align}
\mathscr{M}(\nu, z^2,M_{\pi})=\sum_{k=0}^{k_\text{max}}\lambda_k\tau^k,
\end{align}
where $\tau=\frac{\sqrt{\nu_\text{cut}+\nu}-\sqrt{\nu_\text{cut}}}{\sqrt{\nu_\text{cut}+\nu}+\sqrt{\nu_\text{cut}}}$. Then, we use the fitted $\mathscr{M}(\nu, z^2)$ in the integral in Eq.~\ref{evolution}. The $z$-dependence in the $\mathscr{M}(u\nu, z^2)$ term in the evolution function comes from the one-loop matching term, which is a higher-order correction compared to the tree-level term; thus, the $z$-dependence can be neglected in $\mathscr{M}(\nu, z^2)$.
We choose the dimensionless cutoff $\nu_\text{cut}=1$ as used in the past pseudo PDF calculation~\cite{Joo:2019bzr}. We also vary $\nu_\text{cut}$ between [0.5,2] and the results are consistent with each other. We fix the $\lambda_0=1$ because of the normalization we have for the reduced ITD $\mathscr{M}(\nu, z^2)$ in Eq.~\ref{eq:RITD}. The maximum term $k_\text{max}=3$ is used, because we can fit all the data points $P_z\in [1,5]$ and $z\in [1,5]\times a$ with small $\chi^2$ using a 4-term $z$-expansion.

\begin{figure*}[htbp]
\centering
	\centering
	\includegraphics[width=0.32\linewidth]{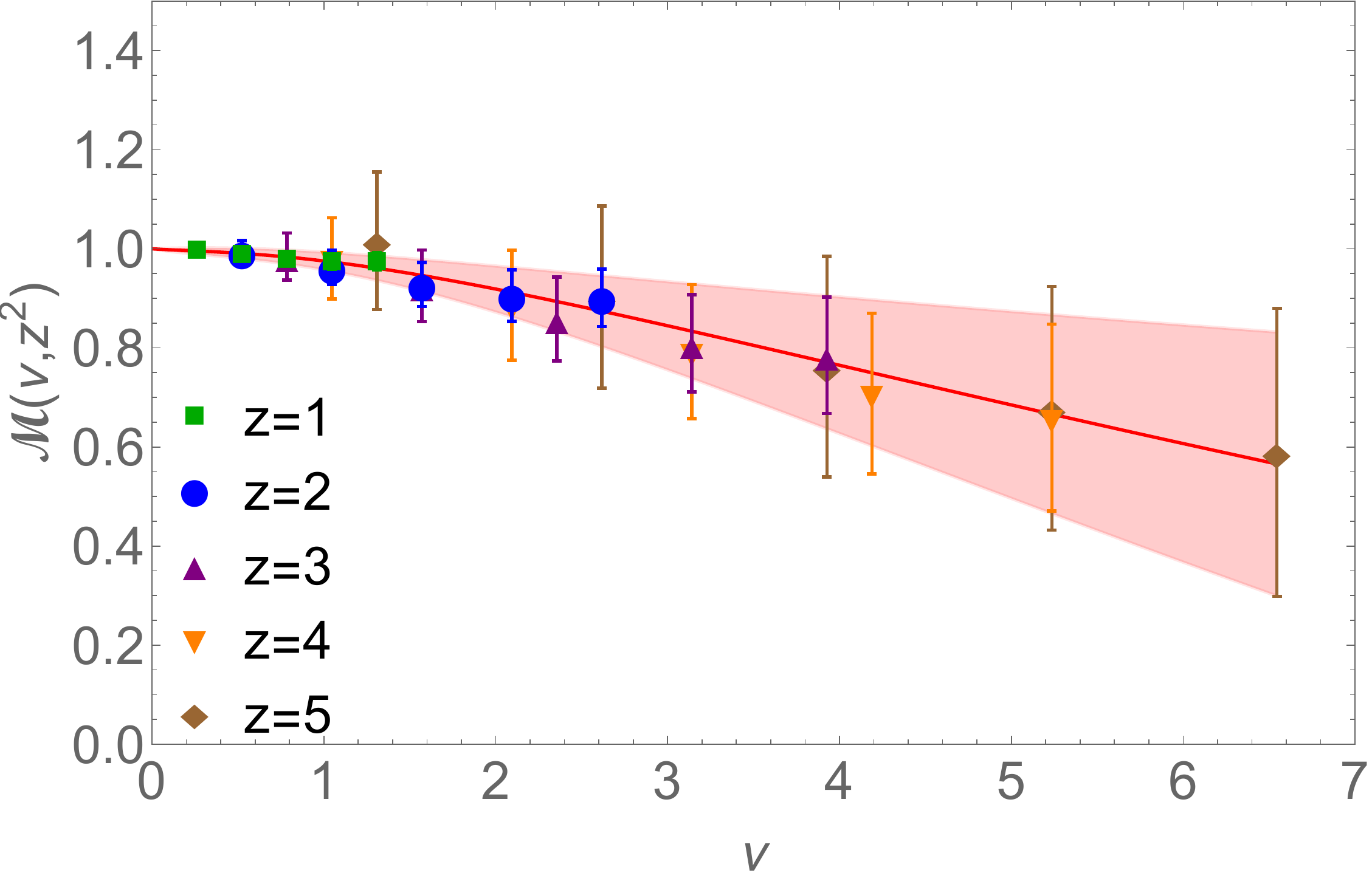}
	\includegraphics[width=0.32\linewidth]{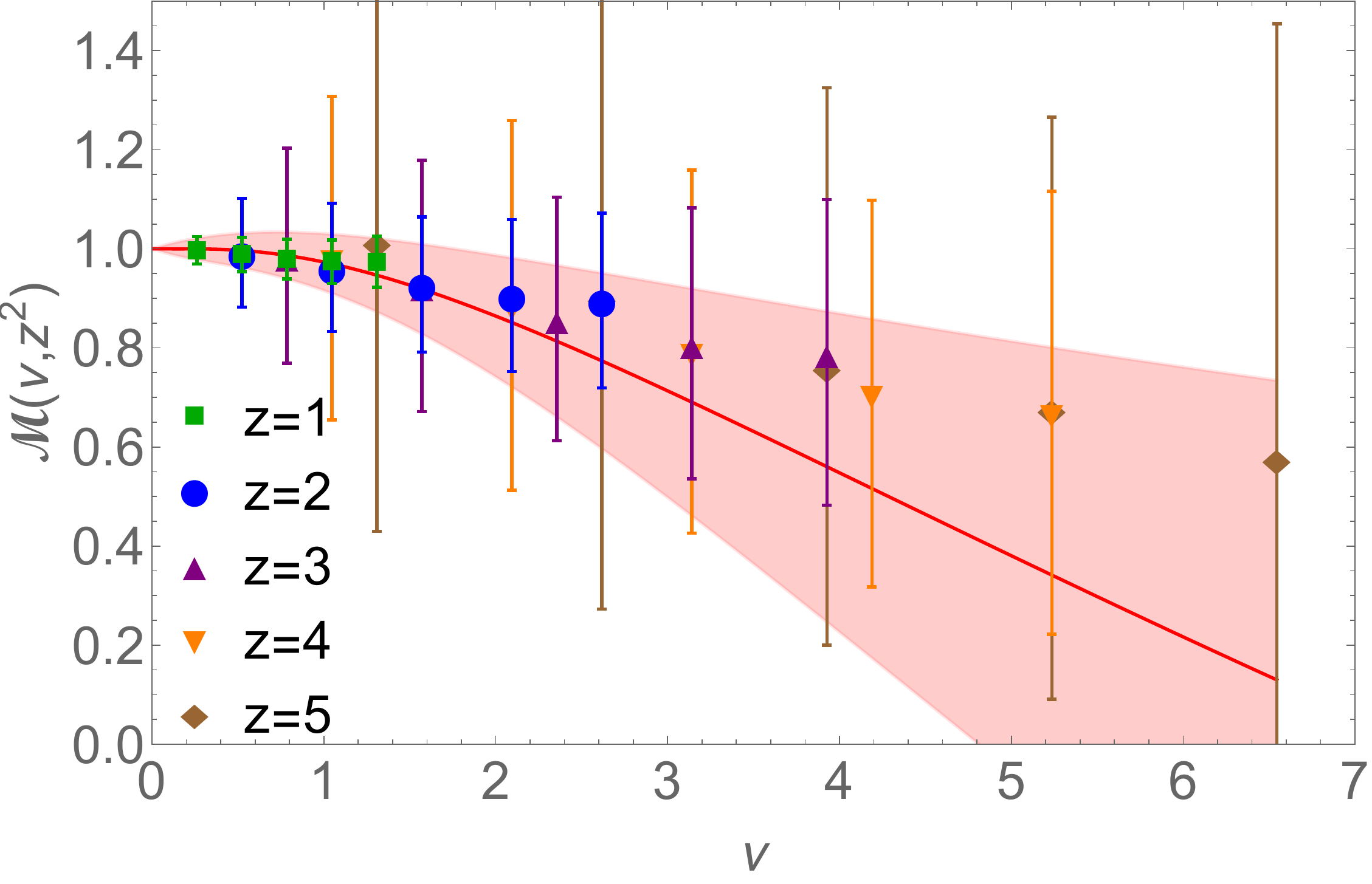}
	\includegraphics[width=0.32\linewidth]{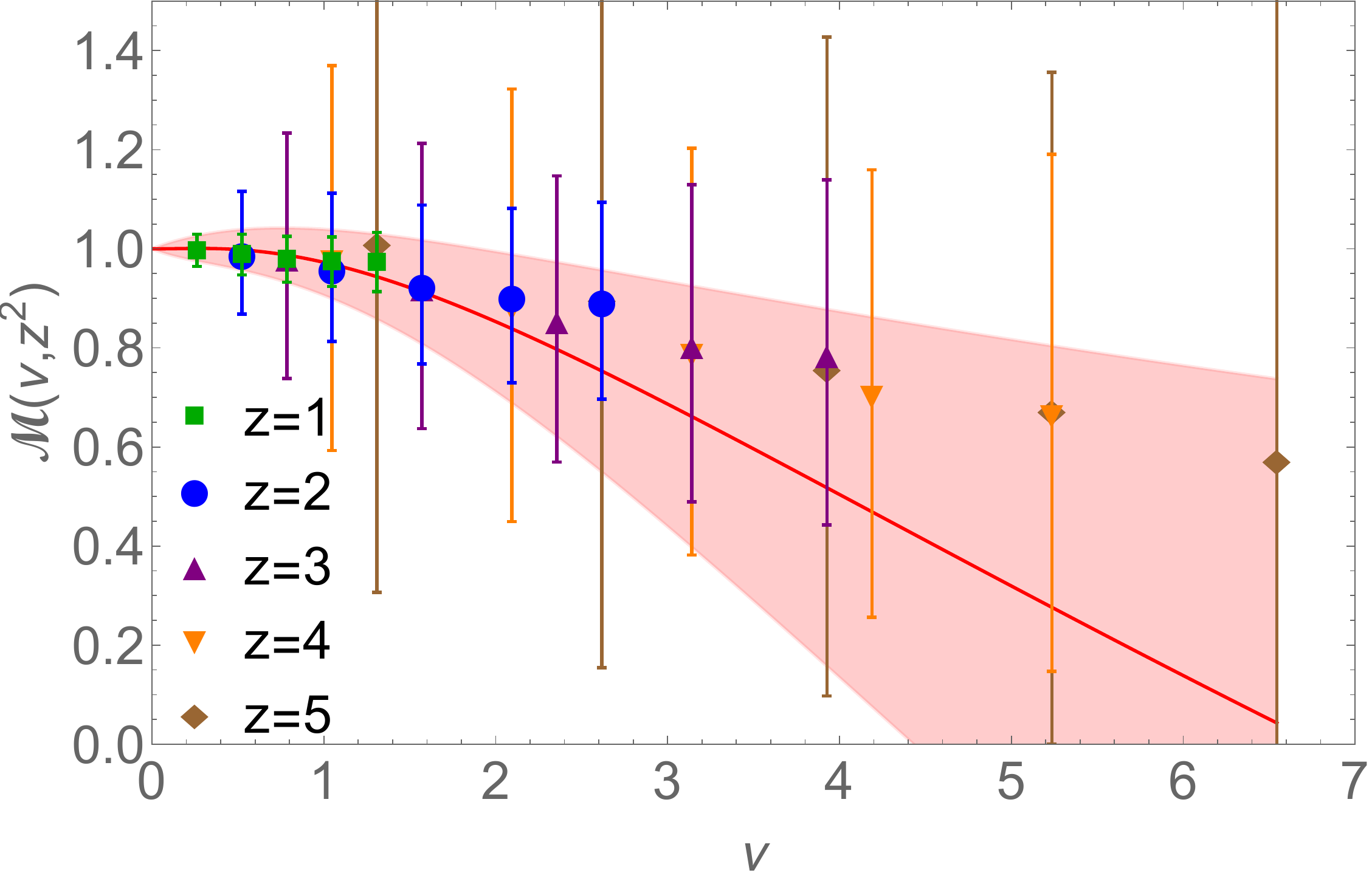}
\caption{The reduced ITDs $\mathscr{M}(\nu, z^2)$ as functions of $\nu$ at pion masses $M_{\pi}=690$, 310 and extrapolated 135~MeV from left to right, respectively. The points of different colors represent the reduced ITDs $\mathscr{M}(\nu, z^2)$ of different $z^2$ and the red band represents the $z$-expansion fit band.
}
\label{fig:ITDs}
\end{figure*}

As shown in Fig.~\ref{fig:ITDs}, the reduced ITDs of different $z^2$ from our lattice calculation show very little $z$ dependence, because the $z$ dependence cancels out when dividing out the ITD at $P=0$ in the ratio defining the reduced ITD. Our fitted bands from the $z$-expansion fit match the reduced ITDs at different pion masses within the error bands. In Fig.~\ref{fig:ITDs}, we can see that the fitted bands are mostly controlled by the small-$z$ reduced ITDs, because the error grows significantly with increasing $z$. The reduced ITDs at physical pion mass are extrapolated from the pion masses at $M_{\pi}=690$ and 310~MeV and are closer to the smaller pion mass at $M_{\pi}=310$~MeV. As $\nu$ grows, the reduced ITDs decrease from $\mathcal{M}(0,z^2)=1$. The decrease becomes faster when we go to smaller pion masses, but this trend is slight because the pion-mass dependence is weak in our case, as seen in Fig.~\ref{fig:ITDs}, where the data and the fitted bands from 3 different pion masses are consistent within one sigma error.

\begin{figure*}[htbp]
\centering
	\centering
	\includegraphics[width=0.32\linewidth]{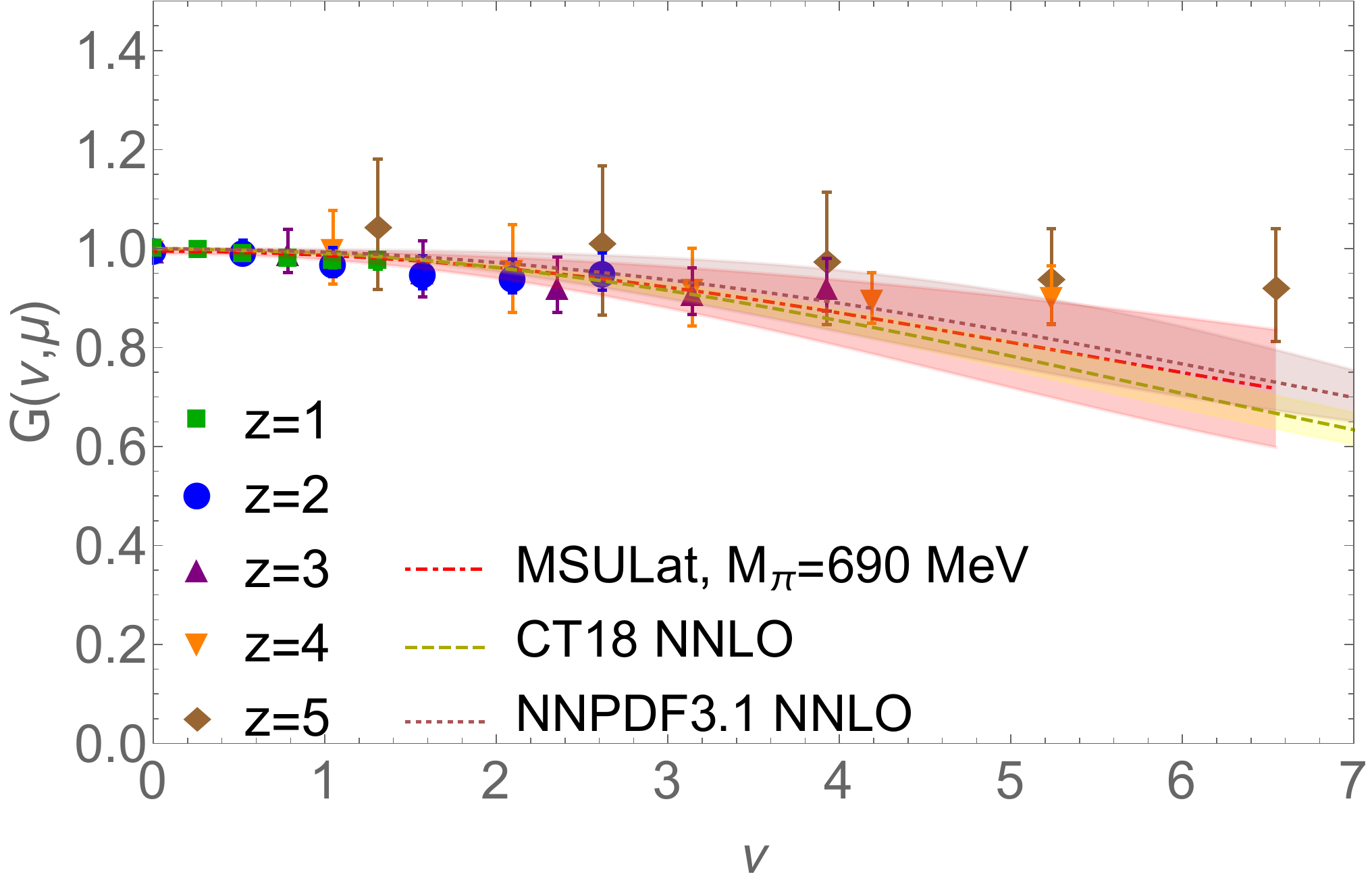}
	\includegraphics[width=0.32\linewidth]{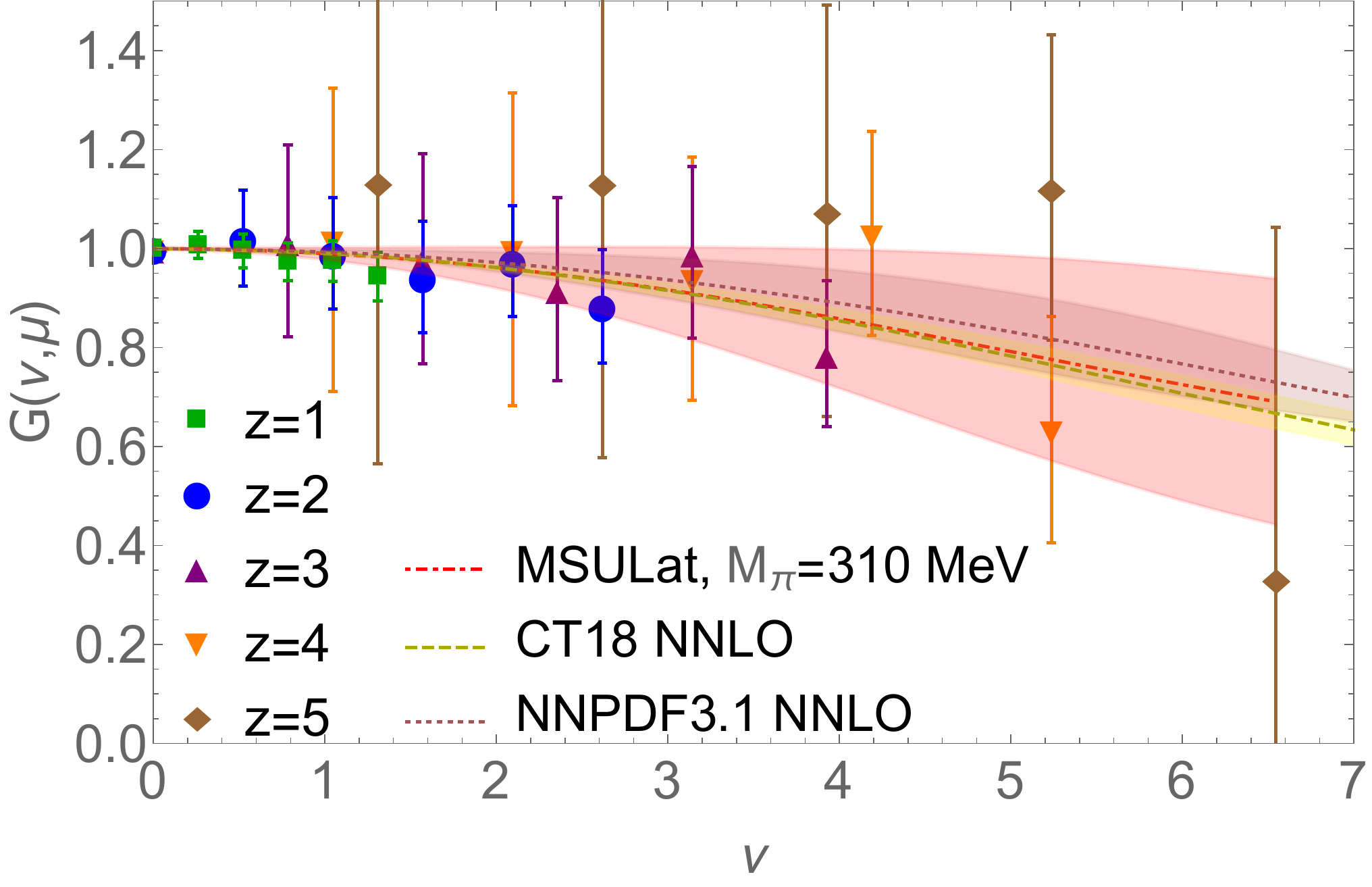}
	\includegraphics[width=0.32\linewidth]{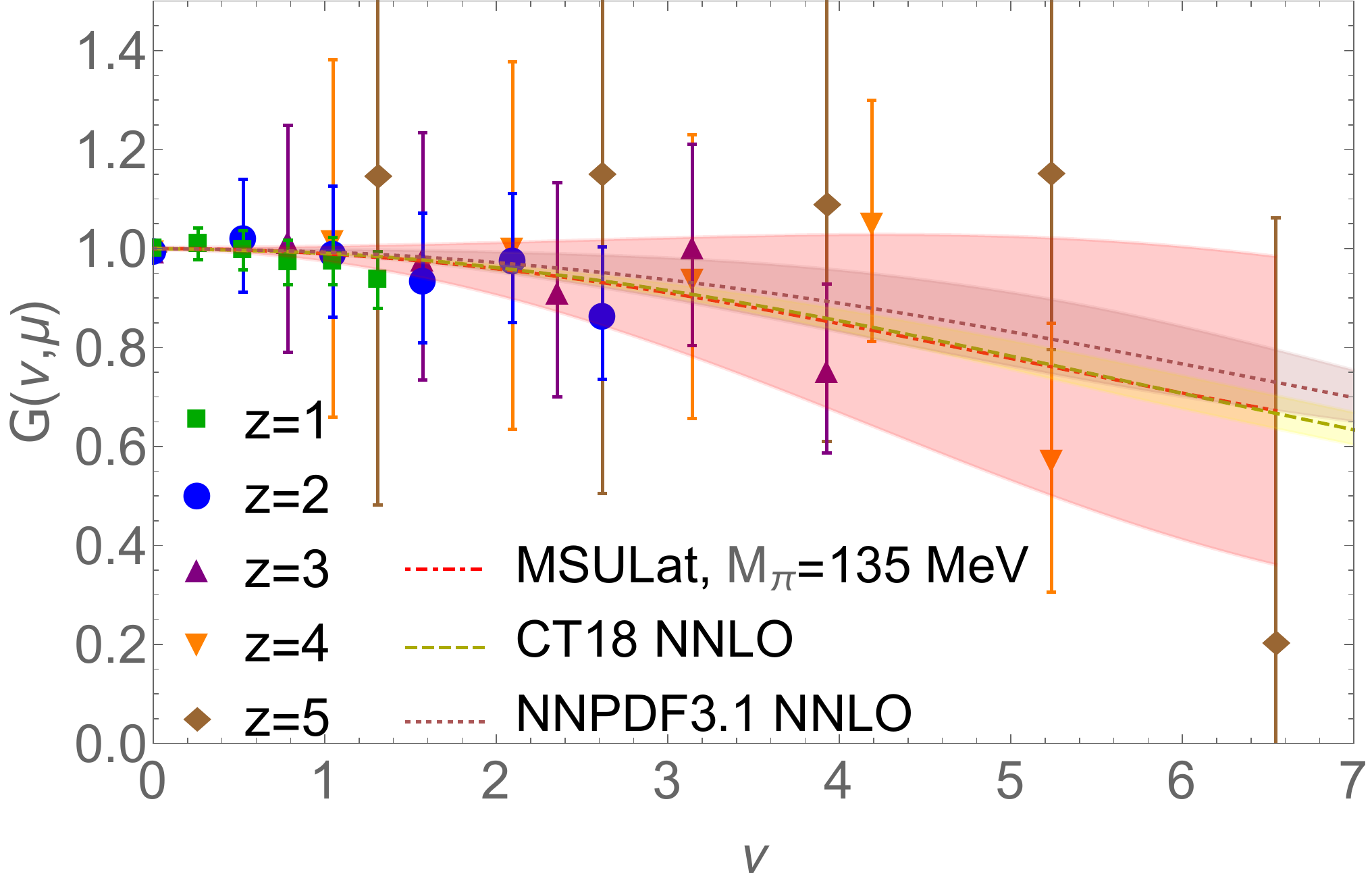}
\caption{The evolved ITDs $G$ as functions of $\nu$ at pion masses $M_\pi \approx 690$, 310 and extrapolated 135~MeV from left to right, respectively. The points of different colors represent the evolved ITDs $G(\nu, z^2)$ of different $z$ values. The red band represents the fitted band of evolved ITD matched from the functional form PDF using the matching formula Eq.~\ref{matching-EITD}. The yellow and pink bands represent the evolved ITD matched from the CT18 NNLO and NNPDF3.1 NNLO unpolarized gluon PDF, respectively. The evolution and matching are both performed at $\mu=2$~GeV in the $\overline{\text{MS}}$ scheme.
}
\label{fig:EITDs}
\end{figure*}

The evolved ITDs at $M_{\pi}=690$, 310 and extrapolated 135~MeV are obtained from Eq.~\ref{evolution}. In the evolution, we choose $\mu=2$~GeV and $\alpha_s(2\text{ GeV})=0.304$. The $z$ dependence of the evolved ITDs should be compensated by the $\ln{z^2}$ term in the evolution formula, which is confirmed in our evolution results. The evolved ITDs from different $z\in [1,5]\times a$ are shown in Fig.~\ref{fig:EITDs} as points with different colors and are consistent with each other within one sigma error. Similar to the reduced ITDs, the evolved ITDs show small pion-mass dependence, because the data points from 3 different pion mass are consistent within one sigma error. According to the evolution function in Eq.~\ref{matching-evolve}, we can obtain the evolved ITD $G$ by adding the reduced ITD $\mathscr{M}$ and an integral term related to $\mathscr{M}$. Due to the cancellation between the two terms, this can reduce the error in the evolved ITDs. This phenomenon is also seen in other pseudo-PDF calculations~\cite{Joo:2019jct,Bhat:2020ktg}.

We assume a functional form for the lightcone PDF to fit the evolved ITD,
\begin{equation}
f_g(x,\mu)=\frac{xg(x, \mu)}{\langle x_g \rangle_{\mu^2}} =\frac{x^A(1-x)^C}{B(A+1,C+1)},
\label{functional}
\end{equation}
for $x\in[0,1]$ and zero elsewhere. The beta function $B(A+1,C+1)=\int_0^1 dx\, x^A(1-x)^C$ is used to normalize the area to unity. The $xg(x, \mu)$ can be reconstructed by multiplying the gluon momentum fraction $\langle x_g \rangle_{\mu^2}=0.411(8)$~\cite{Lin:2020rut} back to the fit form. Then, we apply the matching formula to obtain the evolved ITD,
\begin{equation}
{G}(\nu,\mu)=\int_0^1 dx\, f_g(x,\mu) R_2(x\nu).
\label{matching-EITD}
\end{equation}
We then fit the evolved ITD ${G}$ from the functional form PDF to the evolved ITD $G$ from the lattice calculation. The fits are performed by minimizing the $\chi^2$ function,
\begin{equation}
\chi^2 = \sum_{\nu,z}\frac{({G}(\nu,\mu)-{G}(\nu,z^2,\mu,M_{\pi}))^2}{\sigma^2_{{G}}(\nu,z^2,\mu,M_{\pi})}.
\end{equation}
%

%\textcolor{red}{As shown in Fig.~\ref{fig:EITDs}, we derived evolved ITDs from the CT18 NNLO gluon unpolarized PDF, and we can use the same fit procedure on them. We fit the evolved ITDs with 3 different maximum $\nu$. The fit parameters and $\chi^2/\text{dof}$ are summarized in Table~\ref{tab:CT18-xgfit}. Figure~\ref{fig:CT18-fitPDF} shows a comparison of the reconstructed unpolarized gluon PDF from the form in Eq.~\ref{functional} with the CT18 NNLO gluon unpolarized PDF at $\mu=2$~GeV in the $\overline{\text{MS}}$ scheme. From the comparison plots, we can see that the fit reconstructs the original PDF well in the large-$x$ region but works poorly in the small-$x$ region with the finite-$\nu$ range used in the fit. There are systematic errors in the small-$x$ region yet to be studied.}

\begin{figure*}[htbp]
\centering
	\centering
	\includegraphics[width=0.32\linewidth]{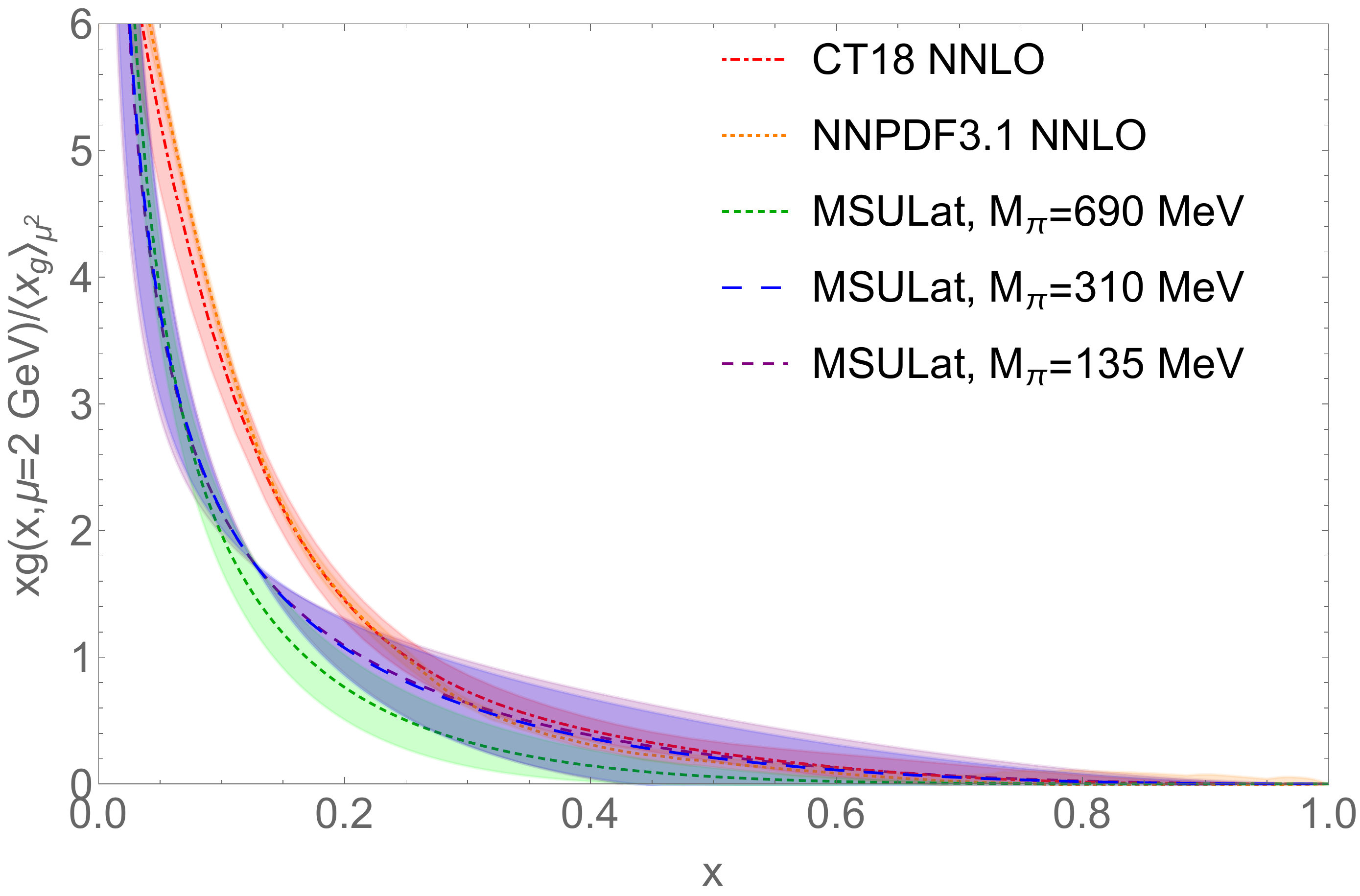}
	\includegraphics[width=0.32\linewidth]{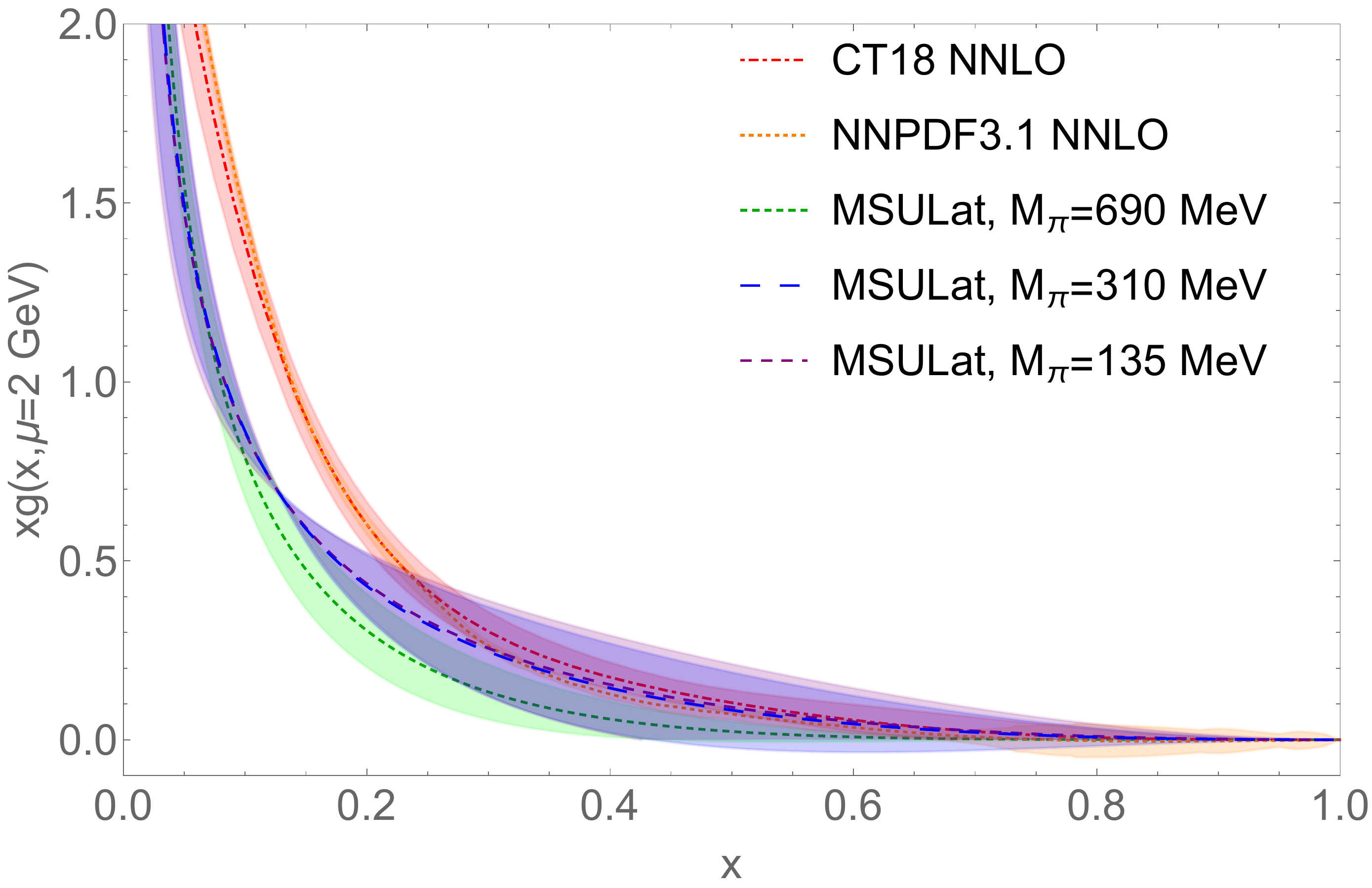}
	\includegraphics[width=0.32\linewidth]{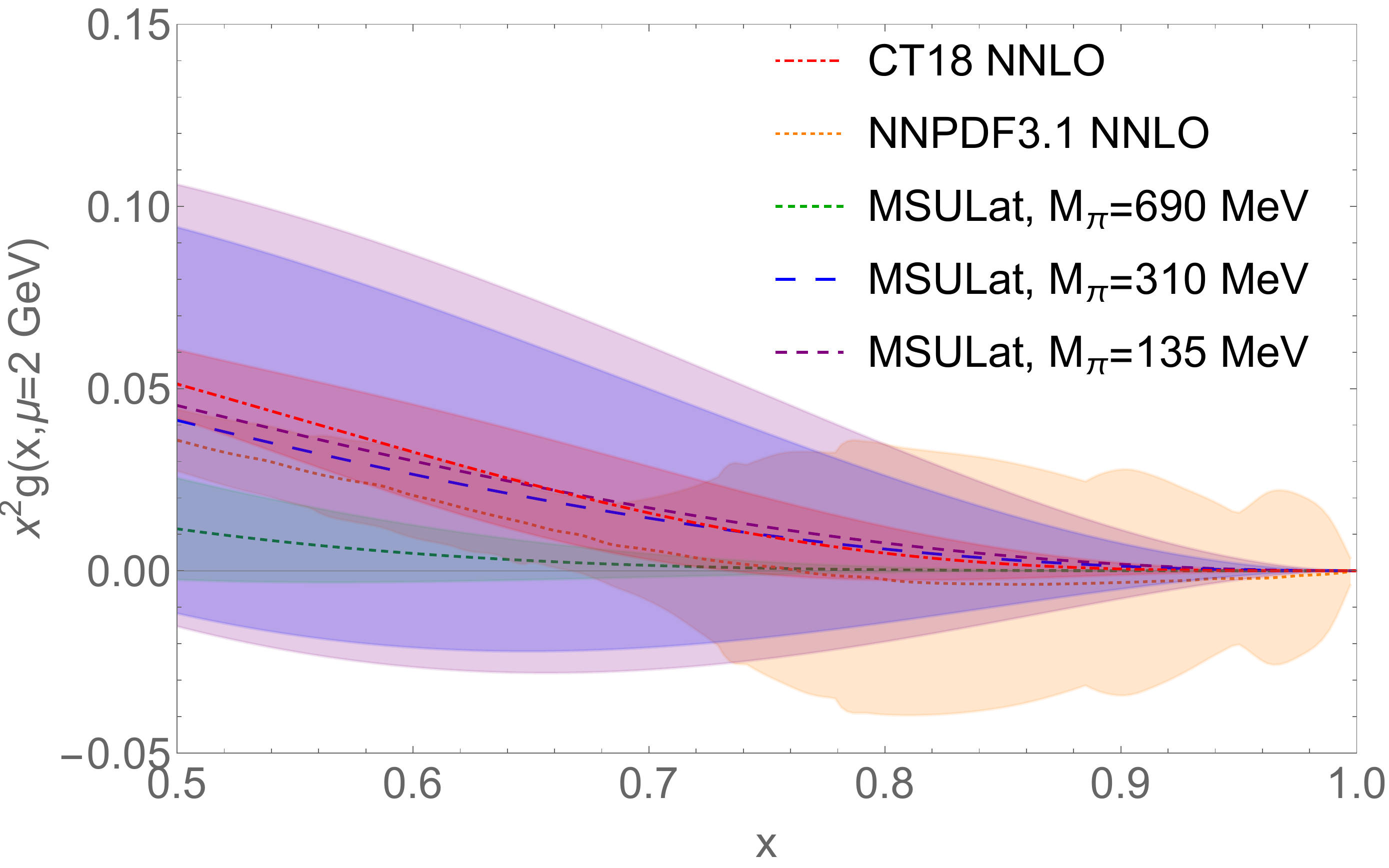}
\caption{
%\FIXME{can you add another left panel for $xg(x)/<x>_g$ plot?}
The unpolarized gluon PDF, $xg(x,\mu)/\langle x_g \rangle_{\mu^2}$ (left), $xg(x,\mu)$ (middle), $x^2g(x,\mu)$ in the large-$x$ region as a function of $x$ (right), obtained from the fit to the lattice data at pion masses $M_\pi=135$ (extrapolated), 310 and 690~MeV compared with the CT18 NNLO (red band with dot-dashed line) and NNPDF3.1 NNLO (orange band with solid line) gluon PDFs.
%The left panel shows $xg(x,\mu)/\langle x_g \rangle_{\mu^2}$ as a function of $x$, middle panel shows $xg(x,\mu)$ as a function of $x$, and the right panel shows $x^2g(x,\mu)$ in the large-$x$ region.
Our $x> 0.3$ PDF results are consistent with the CT18 NNLO and NNPDF3.1 NNLO unpolarized gluon PDFs at $\mu=2$~GeV in the $\overline{\text{MS}}$ scheme.
}
\label{fig:fitPDF}
\end{figure*}

The fit is performed on the evolved ITDs for $M_{\pi}=690$, 310 and extrapolated 135~MeV separately. The fitted evolved ITD represented by the red band shows a decreasing trend as $\nu$ increases. The fit results for three pion masses are consistent with each other, as well as the evolved ITD from CT18 NNLO and NNPDF3.1 NNLO gluon unpolarized PDF, within one sigma error. However, the rate at which it decreases for smaller pion mass is slightly faster. The fit parameters and the goodness of the fit, $\chi^2/\text{dof}$, are summarized in Table~\ref{tab:xgfit}.
From the functional form, it is obvious that parameter $A$ constrains the small-$x$ behaviour and parameter $C$ constrains the large-$x$ behaviour. However, the small-$x$ results obtained from the lattice calculation are not reliable. This is because the Fourier transform of the Ioffe time $\nu$ is related to the region around the inverse of the $x$ and the large-$\nu$ results of evolved ITDs as shown in Fig.~\ref{fig:EITDs} have large error, which leads to poor constraint on the small-$x$ behaviour of $xg(x,\mu)$. In contrast, the large-$x$ behaviour of $xg(x,\mu)$ is constrained well because of the small error in the evolved ITDs in the small-$\nu$ region. Therefore, we have a plot that specifically shows the large-$x$ region of $x^2g(x,\mu)$ in Fig.~\ref{fig:fitPDF}.

\begin{table}[!htbp]
\centering
\begin{tabular}{|c|c|c|c|}
\hline
  $M_\pi$ (MeV)  &   $A$ &   $C$ &  $\chi^2/\text{dof}$\\
  \hline
  690 & $-0.622(14)$  & $2.5(13)$ & $0.35(45)$ \\
  310 & $-0.611(8)$ & $2.3(23)$ & $0.19(36)$ \\
  135 (extrapolated) & $-0.611(9)$ & $2.2(24)$ & $0.19(38)$ \\
%\hline
% CT18 &  $[0.289,0.531]$ &  $[1.872,3.148]$ & N/A \\
\hline
\end{tabular}
\caption{
Our gluon PDF fit parameters, $A$ and $C$, from Eq.~\ref{functional}, and goodness of the fit, $\chi^2/\text{dof}$, for calculations at two valence pion masses and the extrapolated physical pion mass.
%The numbers listed in ``CT18'' are the corresponding fit parameters taken from ``CT18'' and ``CT18Z'' (different global data inputs) from the global fits in Ref.~\cite{Hou:2019efy} at a reference scale equal to charm quark mass.
%Our $A$ exponents are different from the `CT18'' numbers due to the larger noise-of-signal at large $\nu$ region where $A$ constraint is most strong.
%Our $C$ exponents, even though at the slightly higher scale of 2~GeV, are within the range of the ``CT18'' numbers.
}
\label{tab:xgfit}
\end{table}

%\begin{table}[!htbp]
%\centering
%\begin{tabular}{|c|c|c|c|}
%\hline
%  $\nu_\text{max}$  &   $A$ &   $C$ &  $\chi^2/\text{dof}$\\
%  \hline
%  3 & $-0.610(11)$ & $2(14)$ & $0.1(19)$ \\
%  5 & $-0.6110(79)$  & $1.7(70)$ & $0.1(14)$ \\
%  6.54 & $-0.6113(76)$ & $2.0(20)$ & $0.19(36)$ \\
%\hline
% CT18 &  $[-0.7108,-0.469]$ &  $[1.8720,3.1481]$ & N/A \\
%\hline
%\end{tabular}
%\caption{
%}
%\label{tab:xgfit-numax}
%\end{table}

A comparison of our unpolarized gluon PDF with CT18 NNLO and NNPDF3.1 NNLO at $\mu=2$~GeV in the $\overline{\text{MS}}$ scheme is shown in Fig.~\ref{fig:fitPDF}. We compare our $xg(x,\mu)/\langle x_g \rangle_{\mu^2}$ with the phenomenological curves in the left panel. The middle panel shows the same comparison for $xg(x,\mu)$.
%Our fitted $xg(x,\mu)$ for $M_{\pi}=690$, 310 and extrapolated 135~MeV are shown as different color bands. %ZHOUYOU: This only needs to be in the caption. Main text tells us what the figures tell us, not how to read them.
Our $xg(x,\mu)$ extrapolated to the physical pion mass $M_{\pi}=135$~MeV is close to the 310-MeV results and
%ZHOUYOU: (THIS IS WRONG) also agrees with the phenomenological unpolarized gluon PDF better than the 690-MeV results, as expected.
there is only mild pion-mass dependence compared with the 690-MeV results.
We found that our gluon PDF is consistent with the one from CT18 NNLO and NNPDF3.1 NNLO within one sigma in the $x>0.3$ region.
However, in the small-$x$ region ($x< 0.3$), there is a strong deviation between our lattice results and the global fits. This is likely due to the fact that the largest $\nu$ used in this calculation is less than 7, and the errors in large-$\nu$ data increase quickly as $\nu$ increases.
%As a result, there are no good constraints on the small-$x$ region in the fitting procedure.
To better see the large-$x$ behavior, we multiply an additional $x$ factor into the fitted $xg(x,\mu)$ and zoom into the range $x\in[0.5,1]$ in the rightmost plot of Fig.~\ref{fig:fitPDF}. Our large-$x$ results are consistent with global fits over $x \in [0.5,1]$ though with larger errorbars, except for $x \in [0.9,1]$ where our error is smaller than NNPDF, likely due to using fewer parameters in the fit.
With improved calculation and systematics in the future, lattice gluon PDFs can show promising results.
%There is a point near $x=0.18$ where the $xg(x,\mu)$ has very small error due to the PDF functional form used in the fits which is also seen in other quark pseudo-PDF calculations~\cite{Joo:2019jct,Joo:2019bzr}.
% ZHOUYOU: why is the following sentence here?
%In future work, finer lattice-spacing ensembles with larger boost momenta will be critical to study the small-$x$ dependence from lattice-QCD calculations.

\begin{figure*}[htbp]
\centering
	\centering
	\includegraphics[width=0.32\linewidth]{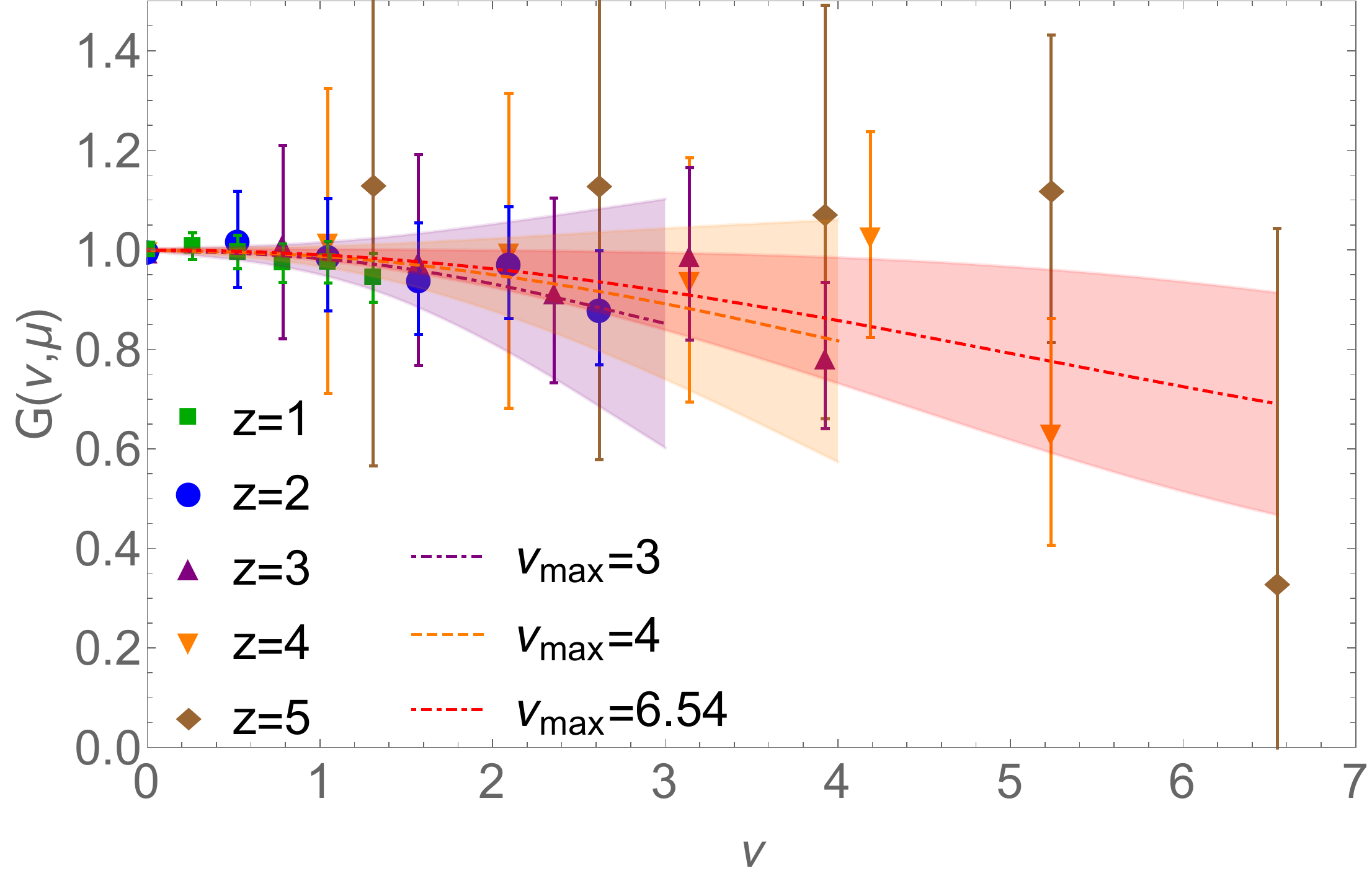}
	\includegraphics[width=0.32\linewidth]{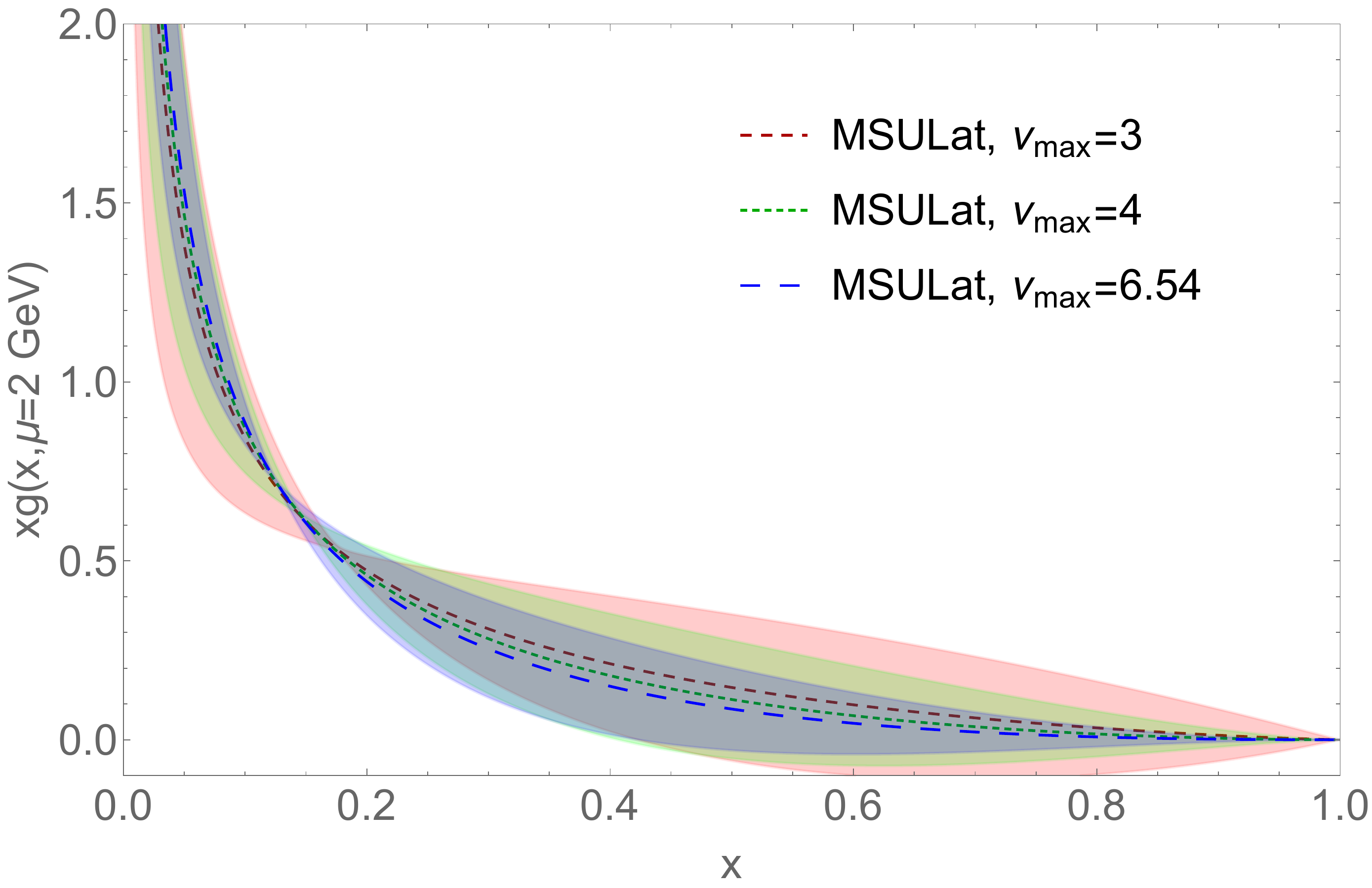}
    \includegraphics[width=0.32\linewidth]{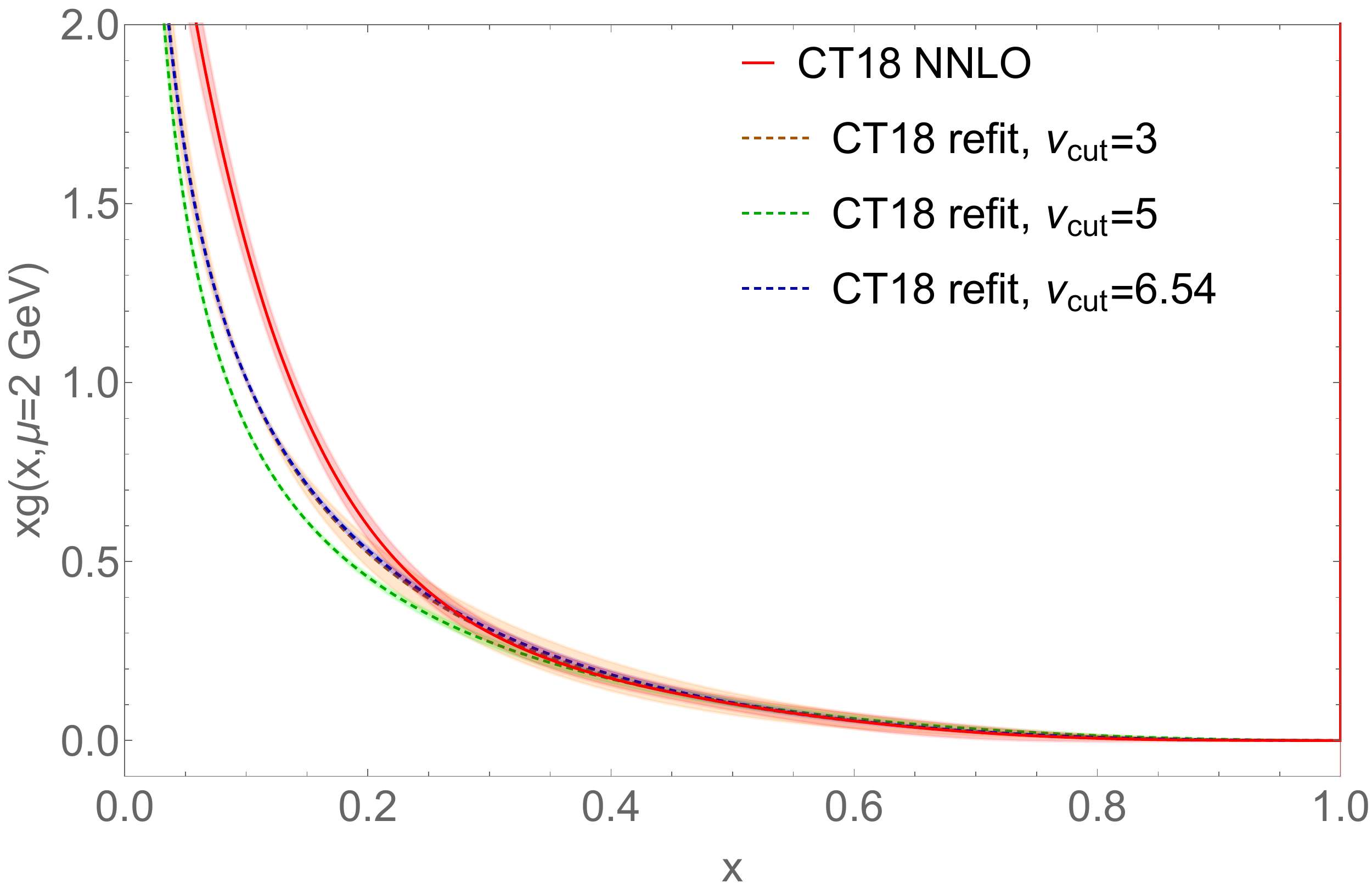}
\caption{Left: The evolved ITDs $G$ as functions of $\nu$ at $M_\pi \approx 310$~MeV with fits performed using different $\nu_\text{max}$ cutoff in the evolved ITDs.
% ZHOUYOU: the following does NOT match the figure descriptions
%The fitted band of the evolved ITD (red band) matching from the functional form PDF from the matching formula Eq.~\ref{matching-EITD}.
%is compared with the evolved ITD matching from the CT18 NNLO (yellow band) and NNPDF3.1 NNLO (pink band) unpolarized gluon PDF respectively.
As we can see from the tightening of the fit band, the evolved ITDs at larger $\nu$ are still useful in constraining the fit despite their larger errors.
Middle: The unpolarized gluon PDF obtained from the fits to the evolved ITDs at 310-MeV pion mass with different $\nu_\text{max}$. %compared with the CT18 NNLO and NNPDF3.1 NNLO unpolarized gluon PDFs is shown in the right plot.
The evolution and matching are both performed at $\mu=2$~GeV in the $\overline{\text{MS}}$ scheme. The larger the $\nu$ input, the more precise the PDF obtained.
Right: The 2-GeV $\overline{\text{MS}}$ renormalized unpolarized gluon PDF obtained from a fit to the evolved ITDs generated from the CT18 NNLO PDF with $\nu_\text{max}\in\{3,5,6.54\}$, compared with the original CT18 NNLO unpolarized gluon PDFs.
%The evolution and matching are both performed at $\mu=2$~GeV in the $\overline{\text{MS}}$ scheme.
As $\nu$ increases, we can see the gluon PDF is better reproduced toward small $x$. Using this exercise, we can see that our lattice PDF is only reliable in the $x>0.25$ region. By taking the moments obtained from CT18 with a cutoff of  $\nu_\text{max}=6.54$ compared to those from the original PDF, we can estimate the higher-moment systematics in our lattice calculation.
}
\label{fig:numax}
\end{figure*}

To demonstrate the influence of the large-$\nu$ data on the fit results, we perform fits to the evolved ITDs with $\nu_\text{max}$ of 3 and 4, comparing with the original fits with $\nu_\text{max}=6.54$. The fits with the $\nu_\text{max}$ cutoff are implemented on the lattice-calculated evolved ITDs and the evolved ITDs created by matching the CT18 NNLO gluon PDF. We show the evolved ITDs from the $M_{\pi}=310$~MeV lattice data and the fitted bands on the left-hand side of Fig.~\ref{fig:numax}. % are shown in the left panel.
The errors of the fit bands become smaller as larger-$\nu_\text{max}$ data are included even though the errors in the input points increases. %
As a result, we can see in the middle of Fig.~\ref{fig:numax} that the lattice gluon PDF errors shrink when the large-$\nu$ data help to constrain the fit.

Since our ability to accurately determine the PDFs in the small-$x$ region is limited by the $\nu_\text{max}$ calculated on the lattice,
we study the effect of the $\nu$ cutoff on our obtained $x$-dependent gluon PDF.
To do so, we took the CT18 NNLO gluon PDF to construct a set of evolved ITDs using the same cutoffs $\nu_\text{max}=\{3,4,6.54\}$ used on the 310-MeV PDF.
%compare the $xg(x)$ as function of $x$ at $\mu=2$ GeV plots from the fits of lattice calculated evolved ITDs and CT18 NNLO gluon PDF evolved ITDs. Both fits are chosen the same cutoff of $\nu_\text{max}=\{3,4,6.54\}$.
The right-hand side of Fig.~\ref{fig:numax} shows that when $\nu_\text{max}$ increases, the region the reconstructed PDF can recover extends to smaller $x$. Based on this observation, we estimate that with $\nu_\text{max}=6.54$, the smallest $x$ at which our lattice PDF can be trusted is around 0.25.
We use the difference between the original CT18 input and the one reconstructed with a $\nu$ cutoff to estimate the systematic due to this cutoff effect on the higher moments.
%We find that the fit reconstructs the original PDF well in the large-$x$ region but works poorly in the small-$x$ region with the finite-$\nu$ range used in the fit with the comparison of refitted CT18 NNLO gluon PDF bands of different $\nu_\text{max}$ and the original CT18 NNLO gluon PDF band, as shown in the right panel of Fig.~\ref{fig:numax}. Therefore, we know that our fitted bands from our lattice data, shown in the middle panel of Fig.~\ref{fig:numax}, should perform well at large-x region and have unknown systematic errors in the small-$x$ region yet to be studied.

 We summarize our predictions for the second and third moments $\langle x_g^2 \rangle_{\mu^2}$ and $\langle x_g^3 \rangle_{\mu^2}$ at $\mu=2$~GeV with their statistical and systematic errors in Table~\ref{tab:moments}, together with the ones from CT18 NNLO and NNPDF3.1 NNLO results. The first error on our number corresponds to the statistical errors from the calculation, while the second error comes from combining in quadrature the systematic errors from four different sources:
1) The normalization of the global-PDF determination of the moment used in our calculation;
2) The finite-$\nu$ cutoff in the evolved ITDs, as discussed above.
 %To study the systematic error coming from the finite-$\nu$ range in our lattice data, we refit the evolved ITDs from the CT18 NNLO gluon unpolarized PDF using the functional form in in Eq.~\ref{functional} with $\nu_\text{max}=3$, $\nu_\text{max}=4$, and $\nu_\text{max}=6.54$ which is the largest $\nu$ in our lattice data. We use the CT18 NNLO results from the CT18 website~\cite{CT18home} and do a naive uncorrelated Gaussian sampling on the CT18 data for error estimation. The moments obtained from the functional form fit with a larger $\nu_\text{max}$ become closer to the original CT18 moments. Therefore, we take half the difference between the mean values of the moment results from $\nu_\text{max}=6.54$ and the  original CT18 moments as an estimate of the systematic error in PDF fit with finite-$\nu$ cutoff in the evolved ITDs.
3) The choice of strong coupling constant. To estimate this error, we vary $\alpha_s$ by 10\%. Like previous pseudo-PDF studies~\cite{Joo:2019bzr}, we find that the changes are no more than 5\%;
%\FIXME{What's actually being done in the mixing?}
4) The mixing with the quark singlet sector. We implement the gluon pseudo-PDF full matching kernel including the quark mixing term on CT18 NNLO unpolarized gluon PDF. The contribution of quark is about $4\%$, which is smaller than systematic errors from other sources. A more precise study of the effects of quark mixing on the unpolarized gluon PDF can be done when we have better control of statistical errors and other systematic errors.
Overall, our moments are in agreement with the global-fit results. Future work including lighter pion masses and finer lattice-spacing ensembles will further help us reduce the systematics in the calculation.

%\begin{table*}[!htbp]
%\centering
%\begin{tabular}{|c|c|c|c|}
%\hline
%  moment  & MSULat ($\nu_\text{max}=3$) & MSULat ($\nu_\text{max}=4$) & MSULat ($\nu_\text{max}=25\pi/12$)  \\
%  \hline
%  $\langle x_g^2 \rangle_{\mu^2}$  &  0.067(71) &  0.055(47)  &  0.047(28)    \\
%\hlinehttps://www.overleaf.com/project/5f0322e0910b250001df2ef6
%  $\langle x_g^3 \rangle_{\mu^2}$  & 0.027(49)  &  0.020(29)  & 0.015(16)      \\
%\hline
%\end{tabular}
%\caption{Higher gluon moments from this work at $M_{\pi}=310$~MeV with different $\nu_\text{max}$ evolved ITDs used in the fits.}
%\label{tab:moments-nmax}
%\end{table*}

\begin{table*}[!htbp]
\centering
\begin{tabular}{|c|c|c|c|c|c|}
\hline
  moment  & MSULat (690 MeV) & MSULat (310 MeV) & MSULat (extrapolated 135 MeV) &  CT18  &  NNPDF3.1 \\
  \hline
  % FIXME: Truncate all the numbers to the same number of digits
  $\langle x_g^2 \rangle_{\mu^2}$  & 0.040(15)(3)   &  0.043(26)(4)  &  0.045(30)(4) & 0.0552(76) & 0.048(13) \\
\hline
  $\langle x_g^3 \rangle_{\mu^2}$  & 0.011(6)(2)   & 0.013(14)(3)   &  0.014(17)(3) & 0.0154(37) & 0.011(9)  \\
\hline
\end{tabular}
\caption{
Predictions for the higher gluon moments from this work and the corresponding ones obtained from CT18 NNLO and NNPDF3.1 NNLO global fits. The first error in our number corresponds to the statistical errors from the calculation and the second errors are the systematic errors.
%1) the global-fit determination of the first moment used to normalize our calculation;
%2) the finite-$\nu$ cutoff, which we estimate from the refit of the evolved ITDs from the CT18 NNLO unpolarized gluon PDF;
%3) the choice of strong coupling constant,
%4) mixing with the quark singlet sector.
}
\label{tab:moments}
\end{table*}

%\begin{table*}[!htbp]
%\centering
%\begin{tabular}{|c|c|c|c|c|c|}
%\hline
%  moment  & MSULat (690 MeV) & MSULat (310 MeV) & MSULat (extrapolated 135 MeV) &  CT18  &  NNPDF3.1 \\
%  \hline
  % FIXME: Truncate all the numbers to the same number of digits
%  $\langle x_g^2 \rangle_{\mu^2}$  & 0.0259(97)(5)(25)(8)   &  0.047(28)(1)(3)(1)  &  0.050(33)(1)(3)(1) & 0.0552(76) & 0.048(13) \\
%\hline
%  $\langle x_g^3 \rangle_{\mu^2}$  & 0.0053(35)(10)(5)(5)   & 0.015(16)(3)(1)(1)   &  0.016(19)(3)(1)(1) & 0.0154(37) & 0.011(9)  \\
%\hline
%\end{tabular}
%\caption{
%Predictions for the higher gluon moments from this work and the corresponding ones obtained from CT18 NNLO and NNPDF3.1 NNLO global fits. The first error in our number corresponds to the statistical errors from the calculation.
%The remaining errors are the systematics from these sources: 1) the global-fit determination of the first moment used to normalize our calculation;
%2) the finite-$\nu$ cutoff, which we estimate from the refit of the evolved ITDs from the CT18 NNLO unpolarized gluon PDF;
%3) the choice of strong coupling constant.
%}
%\label{tab:moments}
%\end{table*}

%%%%%%%%%%%%%%%%%%%%%%%%%%%%%%%%%%%%%%%%%%%%%%%%%%%%%%%%%%%%%%%%%%%%%%%%%%%%%%%%
\section{Summary and Outlook}
\label{sec:summary}

In this paper, we present the first lattice calculation of the gluon parton distribution function using the pseudo-PDF method. The current calculation is only done on one ensemble with lattice spacing of 0.12~fm and two valence-quark masses, corresponding to pion masses around 310 and 690~MeV. In contrast to the prior lattice gluon calculation~\cite{Fan:2018dxu}, we now use an improved gluon operator that is proved to be multiplicatively renormalizable. The gluon nucleon matrix elements were obtained using two-state fits. The use of the improved sources in the nucleon two-point correlators allowed us to reach higher nucleon boost momentum.  As a result, we were able to attempt to extract the gluon PDF as a function of Bjorken-$x$ for the first time.
There are systematics yet to be studied in this work. Future work is planned to study additional ensembles at different lattice spacings so that we can include the lattice-discretization systematics. Lighter quark masses should be used to control the chiral extrapolation to obtain more reliable results at physical pion mass.

%%%%%%%%%%%%%%%%%%%%%%%%%%%%%%%%%%%%%%%%%%%%%%%%%%%%%%%%%%%%%%%%%%%%%%%%%%%%%%%%
\section*{Acknowledgments}

We thank MILC Collaboration for sharing the lattices used to perform this study. The LQCD calculations were performed using the Chroma software suite~\cite{Edwards:2004sx}.
We thank Jian-Hui Zhang and Jiunn-Wei Chen for earlier discussions on the gluon quasi-PDF. We thank Yi-Bo Yang and Raza Sufian for helpful comments.
This research used resources of the
National Energy Research Scientific Computing Center, a DOE Office of Science User Facility supported by the Office of Science of the U.S. Department of Energy under Contract No. DE-AC02-05CH11231 through ERCAP;
facilities of the USQCD Collaboration, which are funded by the Office of Science of the U.S. Department of Energy,
and supported in part by Michigan State University through computational resources provided by the Institute for Cyber-Enabled Research (iCER).
ZF, HL and RZ are partly supported by the US National Science Foundation under grant PHY 1653405 ``CAREER: Constraining Parton Distribution Functions for New-Physics Searches''.

%%%%%%%%%%%%%%%%%%%%%%%%%%%%%%%%%%%%%%%%%%%%%%%%%%%%%%%%%%%%%%%%%%%%%%%%%%%%%%%%
%merlin.mbs apsrev4-1.bst 2010-07-25 4.21a (PWD, AO, DPC) hacked
%Control: key (0)
%Control: author (72) initials jnrlst
%Control: editor formatted (1) identically to author
%Control: production of article title (-1) disabled
%Control: page (0) single
%Control: year (1) truncated
%Control: production of eprint (0) enabled
%

\end{document}